\documentclass[twocolumn,prl,aps,twocolumn,longbibliography]{revtex4-1}
\usepackage[latin9]{inputenc}
\usepackage{textcomp}
\usepackage{amsbsy}
\usepackage{amstext}
\usepackage{amssymb}
\usepackage{graphicx}
\usepackage[unicode=true,pdfusetitle,
 bookmarks=false,
 breaklinks=false,pdfborder={0 0 1},backref=false,colorlinks=false]
 {hyperref}
\hypersetup{
 bookmarksnumbered=false,bookmarksopen=false}

\makeatletter

\newcommand{\lyxmathsym}[1]{\ifmmode\begingroup\def\b@ld{bold}
  \text{\ifx\math@version\b@ld\bfseries\fi#1}\endgroup\else#1\fi}

\@ifundefined{textcolor}{}
{%
 \definecolor{BLACK}{gray}{0}
 \definecolor{WHITE}{gray}{1}
 \definecolor{RED}{rgb}{1,0,0}
 \definecolor{GREEN}{rgb}{0,1,0}
 \definecolor{BLUE}{rgb}{0,0,1}
 \definecolor{CYAN}{cmyk}{1,0,0,0}
 \definecolor{MAGENTA}{cmyk}{0,1,0,0}
 \definecolor{YELLOW}{cmyk}{0,0,1,0}
}

\usepackage{amsmath}
\usepackage{graphicx}
\usepackage{amssymb}
\usepackage{txfonts,color}

\makeatother

\begin{document}

\title{Positioning Nuclear Spins in Interacting Clusters for Quantum Technologies and Bio-imaging}

\author{Zhen-Yu Wang}
\email{zhenyu3cn@gmail.com}
\affiliation{Institut f\"ur Theoretische Physik and IQST, Albert-Einstein-Allee 11, Universit\"at
Ulm, D-89069 Ulm, Germany}
\author{Jan F. Haase}
\email{jan.haase@uni-ulm.de}
\affiliation{Institut f\"ur Theoretische Physik and IQST, Albert-Einstein-Allee 11, Universit\"at
Ulm, D-89069 Ulm, Germany}
\author{Jorge Casanova}
\email{jcasanovamar@gmail.com}
\affiliation{Institut f\"ur Theoretische Physik and IQST, Albert-Einstein-Allee 11, Universit\"at
Ulm, D-89069 Ulm, Germany}
\author{Martin B. Plenio}
\email{martin.plenio@uni-ulm.de}
\affiliation{Institut f\"ur Theoretische Physik and IQST, Albert-Einstein-Allee 11, Universit\"at
Ulm, D-89069 Ulm, Germany}

\begin{abstract}
We propose a method to measure the hyperfine vectors between a nitrogen-vacancy (NV)
center and an environment of interacting nuclear spins. Our protocol enables the
generation of tunable electron-nuclear coupling Hamiltonians while suppressing
unwanted inter-nuclear interactions. In this manner, each nucleus can be addressed
and controlled individually thereby permitting the reconstruction of the individual
hyperfine vectors. With this ability the 3D-structure of spin ensembles
and spins in bio-molecules can be identified without the necessity of varying the
direction of applied magnetic fields. We demonstrate examples including the complete
reconstruction of an interacting spin cluster in diamond and 3D imaging of all the
nuclear spins in a bio-molecule.
\end{abstract}
\maketitle

\section{Introduction}
Protocols for achieving quantum control and locating the position (positioning) of individual members of nuclear
spin ensembles form fundamental building blocks of large scale quantum registers, and may
become essential tools for the resolution of the structure and dynamics of individual
bio-molecules. The nitrogen-vacancy (NV) center in diamond represents a promising physical
platform for the realisation of these protocols even under ambient conditions. This is due
to the favourable properties of the electron spin of the NV-center which, in addition to its long
coherence times, include the possibility for its optical initialization and read-out together
with its coherent manipulation by microwave fields ~\cite{doherty2013nitrogen,dobrovitski2013quantum,WuJPW15}.

An ensemble of $^{13}\text{C}$ nuclear spins in diamond located in the vicinity of the NV
center constitutes an ideal candidate  for building a robust quantum memory because of its
extremely long coherence times. These nuclear spins can be detected and polarized by the NV
center \cite{zhao2011atomic,kolkowitz2012sensing,taminiau2012detection,zhao2012sensing,london2013detecting,shi2014sensing,mkhitaryan2015highly,Casanova2015AXY} and, importantly, NV-$^{13}\text{C}$ entangling quantum gates can be implemented
\cite{van2012decoherence,taminiau2014universal,liu2013noise} allowing as an example
for the teleportation of quantum information between a nuclear spin and an electron spin in
distant diamond samples \cite{pfaff2014unconditional}. Furthermore, NV centers can be implanted
close to the diamond surface where they can detect the signal of nuclear spins above
the surface with single spin sensitivity \cite{muller2014nuclear} which  suggests
the possibility to examine the structure and dynamics of bio-molecules by means
of carefully designed protocols \cite{cai2013diamond,kost2014resolving,ajoy2015atomic}.

To achieve individual addressing and control of nuclear spins it is necessary to determine
the different hyperfine interactions between the NV electron and each nucleus. When
inter-nuclear interactions can be neglected, individual nuclear spins with distinct Larmor
frequencies can be detected \cite{kolkowitz2012sensing,taminiau2012detection,zhao2012sensing,london2013detecting,mkhitaryan2015highly,Casanova2015AXY}
and their hyperfine fields can be measured by the variation of the direction of static magnetic
fields \cite{zhao2012sensing,cai2013diamond} or with the assumed ability to selectively polarize
individual spins \cite{laraoui2015imaging}. However, in situations involving spin ensembles with
similar Larmor frequencies, single spin addressing and complete characterization of hyperfine
fields are highly demanding. The situation is even more challenging when the nuclei are closely
spaced and interacting between each other via dipolar coupling. In this context, the detection
of a single two-spin cluster located at a suitable distance from the NV center and the identification
of part of their interactions has been achieved~\cite{zhao2011atomic,shi2014sensing}. However, a
method to identify each spin in a spin cluster and thereby completely characterize the
Hamiltonian of the spin-cluster is, to the best of our knowledge, unknown.

Recently, it has been shown that for distant nuclear spins such that an applied magnetic field
can be much stronger than the hyperfine interactions, one can significantly improve the addressing
of individual nuclear spins \cite{Casanova2015AXY}. Such a strong magnetic field is also required
when nuclear control fields are applied, e.g., for decoupling of nuclear dipolar interactions, since
the nuclear Zeeman energies should be much stronger than the amplitudes of nuclear control fields.
Importantly, however, under such magnetic fields the method for reconstructing the hyperfine vector
described in \cite{zhao2012sensing} is not useful anymore. This is because a magnetic field not
oriented along the NV symmetry axis causes a different spin mixing in the NV electron ground and
excited states. This destroys the optical qualities damaging, therefore, the initialization and
readout of NV center. Furthermore, it has been experimentally shown that tilting the magnetic field
away from the NV axis reduces the collected photoluminescence, and the difference in fluorescence
counts could be noticeable even when the angle between the magnetic field and NV axis is as small
as $0.1^{\circ}$ \cite{Epstein:2005:94}.

\begin{figure}
\includegraphics[width=0.95\columnwidth]{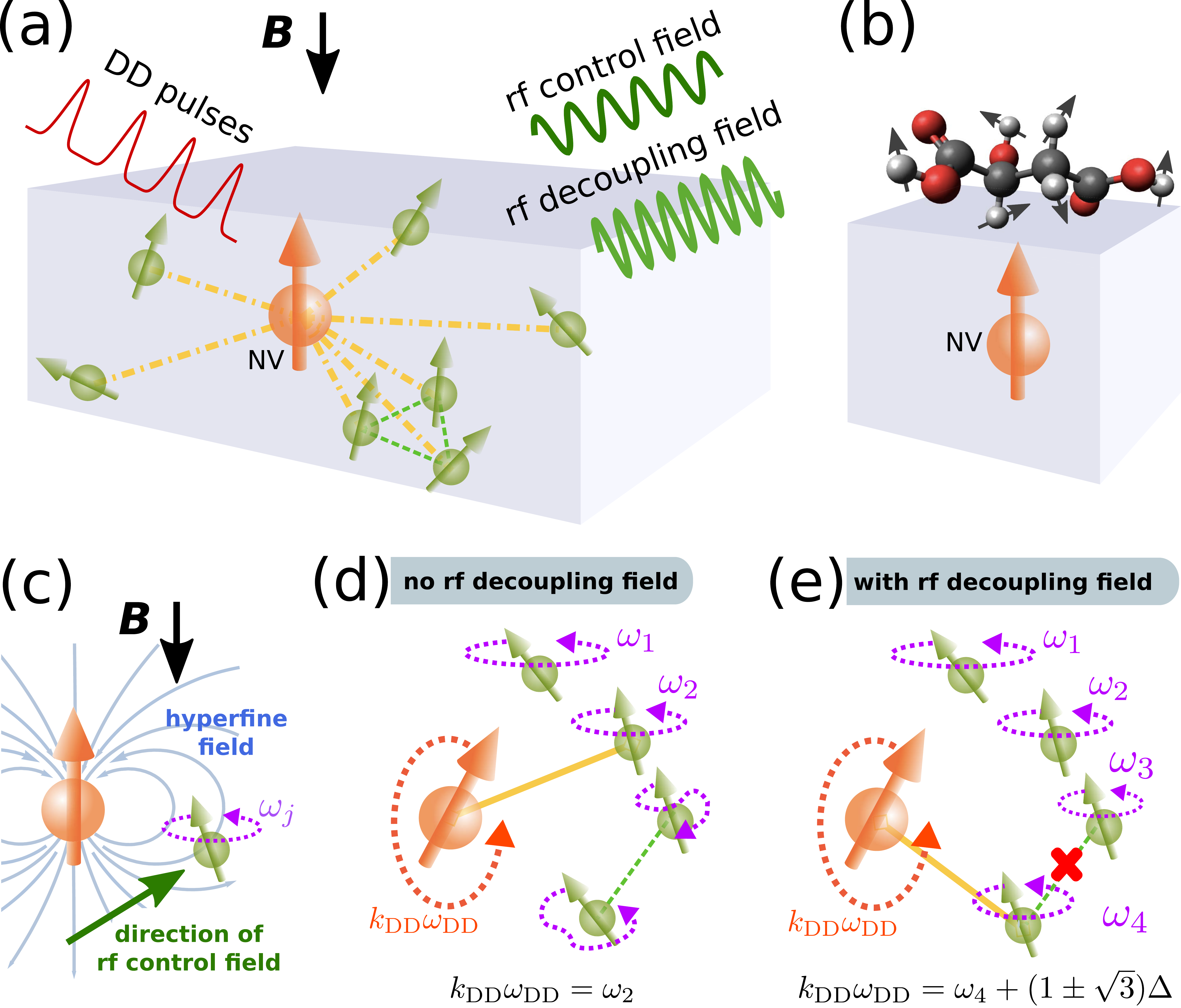}\caption{\label{fig:FigDrawing}(color online).
Individual 3D positioning and control of nuclear spins with an NV center in diamond, by using
rf decoupling and control fields and DD pulse sequences. (a) Our method enables selective addressing
and control of a single nuclear spin (denoted by the yellow dashed dotted lines), e.g., in a coupled
cluster (with nuclear dipolar coupling highlighted by green dashed lines) thus enabling quantum information
processing protocols in quantum registers. (b) Application of the method to structure analysis of a
single molecule by 3D positioning of spins which may be coupled by dipolar interactions.
(c)-(e) Illustrations of the principles of the scheme for measurement and control. (c) The Larmor
frequencies $\omega_{j}$ of nuclear spins are shifted by the position-dependent hyperfine field,
which is symmetric about the NV axis. The 3D positions of nuclear spins are measured by the combination
of the rf control and the hyperfine field, where the former is used to break the symmetry of the system.
(d) When the characteristic flipping frequency of the NV electron spin under DD control matches the Larmor frequency of
the target spin, the target spin is addressed through resonant coupling (denoted by yellow solid line).
Without rf decoupling field, coupled spins (linked by green dashed lines) do not precess at their Larmor frequencies
and can not be individually addressed. (e) The rf decoupling field suppresses the nuclear dipolar
interactions and enables individual addressing and control of the spins in the coupled spin cluster.
}
\end{figure}

In this work, we overcome the difficulties in the measurement of the hyperfine
field in spin ensembles and propose a quantum control method for individual spin
manipulation and three-dimensional (3D) positioning of spin ensembles. The method
combines a robust dynamical decoupling (DD) protocol \cite{Casanova2015AXY} on NV
electron spin with the presence of two radio-frequency (rf) fields acting on the nuclear spins, see
Fig.~\ref{fig:FigDrawing}. One of the rf fields, the rf decoupling field, is used
to eliminate the internuclear interactions while the other one, the rf control field,
is responsible for achieving the selective spin manipulation. The measurement
of the directions of hyperfine fields is achieved by using the phase of rf control
field to break the symmetry of the system Hamiltonian. Remarkably, this protocol
is insensitive to amplitude fluctuations in the rf control field.

Without the need for magnetic field reorientation, which is time consuming in
current laboratory set-ups and also affects fluorescence properties of NV centers, our method can implement fast detection, structure analysis,
and control of spin ensembles. With the ability to work at strong magnetic fields without
changing their directions, good qualities of the optical control on NV electron spin
are preserved, and therefore our method has applications in the detection of homonuclear
spins with chemical shifts which is relevant for structure determination of bio-molecules.

The paper is organized as follows. In Sec.\ref{sec:TheoryMethod} we describe the theory
that is underlying individual addressing and control of nuclear spins in a nuclear
ensemble by an NV center together with the methods to eliminate the internuclear coupling.
In Sec.\ref{sec:Positioning}, schemes to measure the hyperfine interactions and resolving
the structure of nuclear spin ensembles are presented with detailed numerical demonstrations,
which include 3D positioning of a non-interacting spin ensemble, complete characterization
of an interacting $^{13}{\text{C}}$ spin cluster, and 3D positioning and imaging of all the nuclear
spins in a bio-molecule. The conclusions are drawn in Sec.\ref{sec:Summary}.

\section{Theory and Method \label{sec:TheoryMethod}}

\subsection{Microscopic model}
The quantum system composed of an NV center electron spin and its nuclear environment is described by the
Hamiltonian
\begin{equation}
    H=H_{\text{NV}}^{\text{GS}}+H_{\text{hf}}+H_{\text{nZ}}+H_{\text{nn}}.
\end{equation}
Here the ground-state NV electron spin Hamiltonian is  $H_{\text{NV}}^{\text{GS}}=D(\boldsymbol{S}\cdot\hat{z})^{2}-\gamma_{e}\boldsymbol{B}\cdot\boldsymbol{S}$,
where $\boldsymbol{S}=(S_{x},S_{y},S_{z})$ is a spin-1 operator, the axis of the NV-center is
oriented along z-direction with unit vector $\hat{z}$, $\boldsymbol{B}$ denotes an external (static)
magnetic field, $\gamma_{e}$ is the electron spin gyromagnetic ratio, and the ground state zero field
splitting $D=2\pi\times 2.87$ GHz. The nuclear Zeeman Hamiltonian $H_{\text{nZ}} =
-\sum_{j}\gamma_{j}\boldsymbol{B}\cdot\boldsymbol{I}_{j}$, where $\boldsymbol{I}_{j}=(I_{j}^{x},I_{j}^{y},I_{j}^{z})$
and $\gamma_j$ are the $j$-th nuclear spin operator and gyromagnetic ratio respectively. Since tilting
the magnetic field away from the NV axis will introduce errors in measurements, we fix the
magnetic field along the NV symmetry axis $\boldsymbol{B}=B_{z}\hat{z}$.
The electron spin couples to the nuclei via the hyperfine interaction $H_{\text{hf}}$,
and the nuclear magnetic dipole-dipole interaction $H_{\text{nn}}$ reads
\begin{equation}
    H_{\text{nn}}=\sum_{j>k}\frac{\mu_{0}}{4\pi}\frac{\gamma_{j}\gamma_{k}}{r_{j,k}^{3}} \left[\boldsymbol{I}_{j}\cdot\boldsymbol{I}_{k}-\frac{3(\boldsymbol{I}_{j}\cdot\boldsymbol{r}_{j,k})
    (\boldsymbol{r}_{j,k}\cdot\boldsymbol{I}_{k})}{r_{j,k}^{2}}\right],\label{eq:Hnn}
\end{equation}
with $\mu_{0}$ being the vacuum permeability, $\boldsymbol{r}_{j,k}=\boldsymbol{r}_{j}-\boldsymbol{r}_{k}$
the difference between the $k$-th and $j$-th nuclear positions, and $r_{j,k}=|\boldsymbol{r}_{j,k}|$.

Typically the electron-nuclear flip-flop terms in the hyperfine interaction $H_{\text{hf}}$ are suppressed
by the large energy splitting ($\sim$GHz) between the spin triplet ground states $|0\rangle$ and $|m_{s}\rangle$ ($m_{s}=\pm1$)
that exceeds significantly the hyperfine coupling.  In this way, we write the Hamiltonian $H$ in
the rotating frame of $H_{\text{NV}}^{\text{GS}}$ and, when restricted to the manifold of two Zeeman
sublevels $|0\rangle$ and $|m_{s}\rangle$, we find
\begin{equation}
    H_{0,m_{s}}=H_{\text{int}}+ H_{\text{nZ}}^{\text{eff}}+H_{\text{nn}}.
\end{equation}
Here the hyperfine interaction between the NV electron spin and nuclear spins reads
\begin{equation}
    H_{\text{int}}=m_{s}\frac{\sigma_{z}}{2}\sum_{j}\boldsymbol{A}_{j}\cdot\boldsymbol{I}_{j},\label{eq:Hint-0}
\end{equation}
with $\sigma_{z}=|m_{s}\rangle\langle m_{s}|-|0\rangle\langle0|$ and $\boldsymbol{A}_{j}$ the hyperfine
vectors. The nuclear Zeeman Hamiltonian becomes
\begin{equation}
     H_{\text{nZ}}^{\text{eff}}=-\sum_{j}\boldsymbol{\omega}_{j}\cdot\boldsymbol{I}_{j},\label{eq:HnZ}
\end{equation}
where
\begin{equation}
    \boldsymbol{\omega}_{j}\equiv\omega_{j}\hat{\omega}_{j}\equiv\gamma_{j}B_{z}\hat{z}-\frac{m_{s}}{2}\boldsymbol{A}_{j},\label{eq:omegaVec}
\end{equation}
and the Larmor frequency $\omega_{j}=|\boldsymbol{\omega}_{j}|$ is shifted by the hyperfine fields
$\boldsymbol{A}_{j}$. It is the hyperfine vector $\boldsymbol{A}_{j}$ that we wish to
measure accurately, which is important in NV center based quantum information processing and
sensing.

\subsection{Quantum control}

\subsubsection{Decoupling of nuclear magnetic dipole interactions} \label{subsubsec:DecouplingNN}

Magnetic dipolar coupling between nuclei, $H_{\text{nn}}$, interferes with the individual NV
center-nucleus entangling processes and creates difficulties for single-spin measurement.
To show how to suppress the dipolar coupling $H_{\text{nn}}$, we first consider a special case
when the Larmor frequencies $\boldsymbol{\omega}_{j}$ and hence $\omega_j = |\boldsymbol{\omega}_{j}|$
are equal for the coupled spins. Under a strong magnetic field, the single-spin flipping term
in $H_{\text{nn}}$ are suppressed, yielding
\begin{equation}
    H_{\text{nn}}\approx\sum_{j>k}\frac{\mu_{0}}{8\pi}\frac{\gamma_{j}\gamma_{k}}{r_{j,k}^{3}}
    \left[1-\frac{3(\boldsymbol{r}_{j,k}\cdot\hat{z})^{2}}{r_{j,k}^{2}}\right]
    \left(3I_{j}^{z}I_{k}^{z}-\boldsymbol{I}_{j}\cdot\boldsymbol{I}_{k}\right).\label{eq:HnnStrongBz}
\end{equation}
The interaction $H_{\text{nn}}$ given by Eq.~(\ref{eq:HnnStrongBz}) can be canceled in a
rotating frame where the spins rotate at an angle of $\arccos(1/\sqrt{3})$ (the ``magic angle'')
with respect to $\boldsymbol{B}$ according to the Lee-Goldburg irradiation method~\cite{lee1965nuclear, bielecki1989frequency, vinogradov2002proton}. This can be achieved by means of
rf-driving with a Rabi frequency $\sqrt{2}\Delta$ and a detuning $\Delta$ with respect to the
nuclear energy $\omega_{j}$~\cite{cai2013large}. When the spinning rate determined by $\Delta$
exceeds the strength of $H_{\text{nn}}$, the internuclear interaction is averaged out.

In the general case, and because of the different values of the hyperfine vectors $\boldsymbol{A}_{j}$
for different nuclei, the Larmor frequencies $\boldsymbol{\omega}_{j}$ have distinct values
and the quality of the decoupling by magic spinning needs to be reassessed. In Appendix
\ref{sec:Appendix} we show that the dipolar coupling $H_{\text{nn}}$ can still be suppressed when
the rf decoupling field is strong compared with the differences of $\boldsymbol{\omega}_{j}$. The
basic idea is the following. The rf decoupling field for suppressing $H_{\text{nn}}$ reads
\begin{equation}
    H_{\text{rfd}}(t) = \sum_{j}\gamma_{j}V_{\text{rfd}}\cos(\omega_{\text{rfd}}t -\phi_{\text{rfd}})\hat{n}_{\text{rf}}\cdot\boldsymbol{I}_{j},\label{eq:Hrfd}
\end{equation}
where the field direction $\hat{n}_{\text{rf}}$ may be fixed by a coplanar waveguide on the
diamond chip. In the interaction picture of the nuclear Zeeman Hamiltonian $H_{\text{nZ}}$,
the parallel component $\boldsymbol{n}_{j}^{z}=(\hat{n}_{\text{rf}}\cdot\hat{\omega}_{j})\hat{\omega}_{j}$
in $H_{\text{rfd}}(t)$ is neglected under the rotating wave approximation (RWA) and $H_{\text{rfd}}(t)$
drives the spins at the Rabi frequencies $\Omega_{j}^{\text{rfd}}=\frac{1}{2}\gamma_{j}V_{\text{rfd}}|
\boldsymbol{n}_{j}(\phi_{\text{rfd}})|$, where the direction of effective rf driving
\begin{equation}
    \boldsymbol{n}_{j}(\phi_{\text{rfd}}) \equiv \boldsymbol{n}_{j}^{x}\cos\phi_{\text{rfd}}+\boldsymbol{n}_{j}^{y}\sin\phi_{\text{rfd}},\label{eq:nj_Phi}
\end{equation}
is perpendicular to $\hat{\omega}_{j}$. Here $\boldsymbol{n}_{j}^{x}=\hat{n}_{\text{rf}}-\boldsymbol{n}_{j}^{z}$
and $\boldsymbol{n}_{j}^{y}=\hat{\omega}_{j}\times\hat{n}_{\text{rf}}$.
Under a strong magnetic field, $\hat{\omega}_{j}\approx\hat{z}$ and
we choose $\Omega_{j}^{\text{rfd}}\approx\frac{1}{2}\gamma_{j}V_{\text{rfd}}\sqrt{1-|\hat{n}_{\text{rf}}\cdot\hat{z}|^{2}}=\sqrt{2}\Delta$
on the target spins. Other spin species with different values for $\gamma_{j}$ are
not driven because of the large frequency mismatches under strong
magnetic fields. For spins with the same $\gamma_{j}$ there is a
shift in the detuning $\Delta$ given by
\begin{equation}
    \delta_{j}\equiv(\omega_{\text{rfd}}-\omega_{j})-\Delta,\label{eq:deltaj}
\end{equation}
which characterizes the shift in $\omega_{j}$. Because the nuclear-nuclear
interactions decay fast with the distance as $1/r_{j,k}^{3}$, only closely
spaced nuclear spins have non-negligible coupling. Those closely spaced spins
have relatively small differences $\delta_{j}$. When the rf driving is much
stronger than $\delta_{j}$, i.e., $\delta_{j}\ll\Delta$, the coupled spins are
following the same magic spinning to a very good approximation and therefore
interactions between them are suppressed either by the magic spinning decoupling
or by a reduction factor $\propto\delta_{j}/\Delta$ (see Appendix \ref{sec:Appendix}).
In this manner,
and for the sake of clarity in the presentation, we neglect $H_{\text{nn}}$ in the
following analytical work but retain it in our numerics.

\subsubsection{DD control on the NV center}
In order to address a single spin, we apply DD pulse sequences, which can be realized by
microwave control or optical driving~\cite{dobrovitski2013quantum,wang2014all}. In this
way decoherence caused by slow magnetic fluctuations (e.g., drifts in static magnetic fields
caused by temperature fluctuations) can be efficiently suppressed by the DD sequences
through the fast averaging of these effects on the spin states~\cite{yang2011preserving}.
Each of the DD $\pi$ pulse flips the electron spin states, i.e., $|0\rangle\leftrightarrow|m_{s}\rangle$,
in a time much shorter than the time scale of other system dynamics.
Therefore in the following the $\pi$ pulses are treated as
instantaneous. Each $\pi$ pulse introduces a minus sign on the $\sigma_{z}$ operator in
$H_{\text{int}}$ {[}Eq.~(\ref{eq:Hint-0}){]}. That is, the DD pulse sequence leads to the
transformation $\sigma_{z}\rightarrow F(t)\sigma_{z}$, where the
modulation function $F(t)$ takes the values $+1$ or $\lyxmathsym{\textminus}1$
depending whether an even or odd number of $\pi$ pulses have been
applied. In this way, under DD control the interaction Hamiltonian reads
\begin{equation}
    H_{\text{int}}(t)=m_{s}F(t)\frac{\sigma_{z}}{2}\sum_{j}\tilde{\boldsymbol{A}}_{j}(t)\cdot\boldsymbol{I}_{j},\label{eq:Hint}
\end{equation}
in a rotating  frame with respect to
\begin{equation}
    H_{\text{n}}(t) =  H_{\text{nZ}}^{\text{eff}} + H_{\text{rfd}}(t),\label{eq:Hn}
\end{equation}
where $ H_{\text{nZ}}^{\text{eff}}$ and $H_{\text{rfd}}(t)$ are given by Eqs.~(\ref{eq:HnZ})
and (\ref{eq:Hrfd}) respectively. In Eq.~(\ref{eq:Hint}) we have $\tilde{\boldsymbol{A}}_{j}(t) \cdot \boldsymbol{I}_{j}=U_{\text{n}}^{\dagger}(t)\boldsymbol{A}_{j}\cdot\boldsymbol{I}_{j}U_{\text{n}}(t)$, where $U_{\text{n}}(t)=\mathcal{T}e^{-i\int_{0}^{t}H_{\text{n}}(\tau)d\tau}$
with $\mathcal{T}$ being the time-ordering operator. We show in Appendix
\ref{sec:Appendix} that
\begin{equation}
\tilde{\boldsymbol{A}}_{j}(t)=\tilde{\boldsymbol{A}}_{j}^{x}(t)\cos(\omega_{\text{rfd}}t)+\tilde{\boldsymbol{A}}_{j}^{y}(t)\sin(\omega_{\text{rfd}}t)+\tilde{\boldsymbol{A}}_{j}^{z}(t),\label{eq:AjTilde}
\end{equation}
\begin{eqnarray}
    \tilde{\boldsymbol{A}}_{j}^{\alpha}(t) & \equiv & (\boldsymbol{A}_{j}^{\alpha}-\boldsymbol{A}_{j}^{\alpha}\cdot\hat{\nu}_{j}\hat{\nu}_{j})\cos(\nu_{j}t)\nonumber \\
    &  & -\hat{\nu}_{j}\times\boldsymbol{A}_{j}^{\alpha}\sin(\nu_{j}t)+\boldsymbol{A}_{j}^{\alpha}\cdot\hat{\nu}_{j}\hat{\nu}_{j},\;\alpha=x,y,z,
\end{eqnarray}
\begin{eqnarray}
    \boldsymbol{A}_{j}^{x} & \equiv & \boldsymbol{A}_{j}-\boldsymbol{A}_{j}^{z},\\
    \boldsymbol{A}_{j}^{y} & \equiv & \hat{\omega}_{j}\times\boldsymbol{A}_{j},\\
    \boldsymbol{A}_{j}^{z} & \equiv & \boldsymbol{A}_{j}\cdot\hat{\omega}_{j}\hat{\omega}_{j},\label{eq:Az}
\end{eqnarray}
under a rf field with $\omega_{\text{rfd}}\sim\omega_{j}\gg\gamma_{j}V_{\text{rfd}}$.
Here $\boldsymbol{A}_{j}^{x}$ and $\boldsymbol{A}_{j}^{y}$ are perpendicular
to $\hat{\omega}_{j}$, and $\hat{\nu}_{j}$ is the unit vector of
\begin{equation}
    \boldsymbol{\nu}_{j}=\frac{1}{2}\gamma_{j}V_{\text{rfd}}\boldsymbol{n}_{j}(\phi_{\text{rfd}})+(\omega_{\text{rfd}}-\omega_{j})\hat{\omega}_{j},\label{eq:nuj}
\end{equation}
whose modulus reads $\nu_{j}=|\boldsymbol{\nu}_{j}|=\sqrt{(\Omega_{j}^{\text{rfd}})^{2}+(\omega_{\text{rfd}}-\omega_{j})^{2}}$.
Physically, $\tilde{\boldsymbol{A}}_{j}(t)$ is the vector $\boldsymbol{A}_{j}$ after two
subsequent rotations. The first rotates the vector $\boldsymbol{A}_{j}$ about
the axis $\hat{\omega}_{j}$ by an angle $\omega_{\text{rfd}}t$ while the second one is
a rotation around the axis $\hat{\nu}_{j}$ by an angle $\nu_{j}t$.

When choosing $\Omega_{j}^{\text{rfd}}=\sqrt{2}\Delta$ for achieving internuclear decoupling, see Appendix \ref{sec:Appendix},
the amplitude of the spinning rate by rf driving
\begin{equation}
\nu_{j}=\Delta\sqrt{2+(1+\delta_{j}/\Delta)^{2}}\approx\sqrt{3}\Delta+\frac{1}{\sqrt{3}}\delta_{j},
\end{equation}
has a strength $\sqrt{3}\Delta$ up to a spin dependent shift. When
there is no rf decoupling field $V_{\text{rfd}}=\omega_{\text{rfd}}=0$,
$\boldsymbol{\nu}_{j}=-\boldsymbol{\omega}_{j}$, and Eq.~(\ref{eq:AjTilde})
reads $\tilde{\boldsymbol{A}}_{j}(t)=\boldsymbol{A}_{j}^{x}\cos(\omega_{j}t)+\boldsymbol{A}_{j}^{y}\sin(\omega_{j}t)+\boldsymbol{A}_{j}^{z}$,
which contains terms that oscillate at the Larmor frequency $\omega_{j}$.
When the rf decoupling field has been applied, $\tilde{\boldsymbol{A}}_{j}(t)$
contains terms that oscillate at the frequencies $\nu_{j}$ and $\omega_{\text{rfd}}\pm\nu_{j}$
which can be obtained by expanding $\tilde{\boldsymbol{A}}_{j}(t)$ with the use
of product-to-sum trigonometric formulas. In both cases the spin-dependent
oscillating harmonics can be utilized to individually address different nuclear spins.

\subsubsection{Individual nuclear spin addressing}\label{sub:addressing}

For periodic DD pulses, the modulation function $F(t)=F(t+\tau_{\text{DD}})$
can be expanded in a Fourier series such that for even $F(t)$ we have
\begin{equation}
    F(t)=\sum_{k\geq1}f_{k}\cos(k\omega_{\text{DD}}t),\label{eq:Ft-Expansion}
\end{equation}
where $\tau_{\text{DD}}=2\pi/\omega_{\text{DD}}$ describes the period of the DD
sequence. The lowest frequency in this expansion, i.e., $\omega_{\text{DD}}$
characterizes the flipping rate of the NV electron spin by the DD pulses. When
$\omega_{\text{DD}}$ or its multiples do not match the characteristic frequencies
of the surrounding nuclear spin ensemble (the Larmor frequencies $\omega_j$ for
the case of no rf decoupling field), then the interaction $H_{\text{int}}(t)$
(see Eq. ~(\ref{eq:Hint-0})) is suppressed. Under DD control we can arbitrarily
tune $\omega_{\text{DD}}$ and when the DD pulses are tuned to a resonance with
the target nuclear spins (e.g., $k_{\text{DD}}\omega_{\text{DD}} = \omega_j$
for the case of no rf decoupling field) the interaction Hamiltonian in Eq.~(\ref{eq:Hint})
couples the NV electron and target nuclear spins

In the literature, the most widely used DD sequences for coherence protection
and spin detection are the Carr-Purcell-Meiboom-Gill (CPMG)~\cite{Carr:1954:630,MeiboomRSI1958}
and the XY family~\cite{MaudsleyJMR1986,GullionJMR1990} sequences of pulses, which
have fixed expansion coefficients $f_{k}=4(k\pi)^{-1}\sin(\frac{k\pi}{2})$ in Eq.~(\ref{eq:Ft-Expansion}).
However, recently it was demonstrated that the ability to tune the expansion
coefficients $f_{k}$ can lead to a considerable enhancement in resolution and
hence single spin addressing~\cite{Zhao2014KDD,AlbrechtP15,Casanova2015AXY,Ma2015resolving}.
In particular, the adaptive-XY (AXY) sequences in Ref. \cite{Casanova2015AXY}
tune the $f_{k}$ for highly selective spin addressing while remaining robust
against detuning and pulse amplitude errors. Under typical error regimes and
for typical control parameters in experiments of DD control on NV centers,
simulations show that the AXY sequences behave well even when the number
of DD pulses reaches several thousands \cite{Casanova2015AXY}. In this work 
we adopt the AXY sequence for its excellent abilities for individual spin
addressing and robustness, even though the theory to be described below is
general and applies to other DD schemes such as CPMG or any decoupling sequence of the XY family.

When the dipolar coupling $H_{\text{nn}}$ is negligible during the time of the
protocol, the rf decoupling field $H_{\text{rfd}}(t)$ in Eq.~(\ref{eq:Hrfd})
is not necessary. If $H_{\text{rfd}}(t)=0$, $\tilde{\boldsymbol{A}}_{j}(t)$
{[}Eq.~(\ref{eq:AjTilde}){]} has components oscillating at the Larmor
frequency $\omega_{j}$. When the frequency of the $k_{\text{DD}}$-th harmonic
in the expansion Eq.~(\ref{eq:Ft-Expansion}) matches the Larmor frequency
$\omega_{n}$ of the $n$-th spin, i.e., the scanning frequency satisfies
\begin{eqnarray}
    \omega_{\text{scan}}=\omega_{n}=k_{\text{DD}}\omega_{\text{DD}},
\end{eqnarray}
we have resonant coupling between the NV center and the $n$-th spin. For a
sufficiently strong magnetic field $\gamma_{j}B_{z}\sim\omega_{j}\gg|\boldsymbol{A}_{j}|$,
we have the effective Hamiltonian under RWA~\cite{Casanova2015AXY}
\begin{equation}
    H_{\text{int}}(t)\approx\frac{m_{s}}{4}f_{k_{\text{DD}}}\sigma_{z}\boldsymbol{A}_{n}^{x}\cdot\boldsymbol{I}_{n},\label{eq:Hint_Ax}
\end{equation}
whose validity requires that the following conditions are satisfied (with $j\neq n$)
\begin{eqnarray}
    |\gamma_{j}B_{z}| & \gg & k_{\text{DD}}|\boldsymbol{A}_{j}|,\label{eq:RWAStrongField}\\
    |\omega_{n}-\omega_{j}| & \gg & |f_{k_{\text{DD}}}\boldsymbol{A}_{j}^{x}|.\label{eq:RWAWeekFdd}
\end{eqnarray}
The first condition, which limits the largest possible $k_{\text{DD}}$, can be reached
by a strong magnetic field while the second condition can be met by the AXY sequences,
which can tune $f_{k_{\text{DD}}}$ to arbitrarily small values without the necessity of
using high harmonics, i.e., high values of $k_{\text{DD}}$. The first condition can also
be reached by a sufficiently large distance between the NV center and the nuclear spins,
because of the reduction of the hyperfine field $\boldsymbol{A}_{j}$. It should be noted
though that a large distance also imposes difficulties on individual spin addressing because
of reduced differences among the hyperfine-field shifted Larmor frequencies $\omega_{j}$.

If an rf decoupling field has been used to suppress
the interaction $H_{\text{nn}}$, $\tilde{\boldsymbol{A}}_{j}(t)$ {[}Eq.~(\ref{eq:AjTilde}){]}
contains terms that oscillate at the frequencies $\nu_{j}$ and $\omega_{\text{rfd}}\pm\nu_{j}$.
When the resonance conditions
\begin{equation}
    k_{\text{DD}}\omega_{\text{DD}}=\nu_{n} \;\;\mbox{or}\;\; k_{\text{DD}}\omega_{\text{DD}}=\omega_{\text{rfd}}\pm\nu_{n} \label{eq:rfRes}
\end{equation}
are satisfied, then there is coupling between the NV electron spin and the $n$-th spin
analogous to Eq.~(\ref{eq:Hint_Ax}). The Larmor frequencies $\omega_n$ can be detected
by scanning simultaneously the frequencies of rf decoupling field and $\omega_{\text{DD}}$ of
pulse sequences. For example, we may choose the rf decoupling field with the Rabi driving
$\Omega_{j}^{\text{rfd}}=\sqrt{2}\Delta$ and frequency
\begin{equation}
    \omega_{\text{rfd}}  =\omega_{\text{scan}}+\Delta,  \label{eq:ResRfd}
\end{equation}
and, at the same time, the DD sequence with
\begin{equation}
    k_{\text{DD}}\omega_{\text{DD}} =\omega_{\text{scan}}+(1\pm\sqrt{3})\Delta. \label{eq:ResWdd}
\end{equation}
When the scanning frequency $\omega_{\text{scan}}=\omega_{n}$, the resonance condition
Eq.~(\ref{eq:rfRes}) is reached with $\delta_{n}=0$ [Eq.~(\ref{eq:deltaj})]. For nearby
spins coupled to the $n$-th spin, $\delta_{j}$ is small, and the rf decoupling field with
$\delta_{j}\ll\Delta$ suppresses the nuclear dipolar coupling to the $n$-th spin. Similar
to the case in obtaining Eq.~(\ref{eq:Hint_Ax}), we can neglect fast oscillating terms
under RWA. The effective Hamiltonian for the resonance Eq.~(\ref{eq:rfRes}) reads
\begin{equation}
    H_{\text{int}}(t)\approx\frac{m_{s}}{8}f_{k_{\text{DD}}}\sigma_{z}\boldsymbol{a}_{n}^{\pm}\cdot\boldsymbol{I}_{n},\label{eq:Hint_RfApm}
\end{equation}
\begin{eqnarray}
    \boldsymbol{a}_{n}^{(\pm)} & \equiv & \boldsymbol{A}_{n}^{x}-\boldsymbol{A}_{n}^{x}\cdot\hat{\nu}_{n}\hat{\nu}_{n}\pm\hat{\nu}_{n}\times\boldsymbol{A}_{n}^{y},\label{eq:an_pm}
\end{eqnarray}
under the conditions Eq.~(\ref{eq:RWAStrongField}),
\begin{eqnarray}
    |\gamma_{j}B_{z}| & \gg & \Delta,\label{eq:RWAStrongBzRF}\\
    \sqrt{3} \Delta & \gg & |f_{k_{\text{DD}}}\boldsymbol{A}_{j}|,\label{eq:RWAStrongRFAj}\\
\left||\gamma_{j}B_{z}|k/k_{\text{DD}} - \sqrt{3}\Delta\right|& \gg &  |\boldsymbol{A}_{j}|, \forall k\geq 1,\label{eq:RWAStrongBzAj}
\end{eqnarray}
and the addressing condition
\begin{equation}
    |\nu_{j}-\nu_{n}|\gg|f_{k_{\text{DD}}}\boldsymbol{A}_{j}|.
\end{equation}
The condition Eq. (\ref{eq:RWAStrongBzRF}) guarantees the validity of rf driving.
The condition Eq. (\ref{eq:RWAStrongField}) eliminates the static terms in $\tilde{\boldsymbol{A}}_{j}(t)$ 
because of the fast oscillation in $F(t)$. The time-dependent terms in $\tilde{\boldsymbol{A}}_{j}(t)$ 
carry the frequencies $\nu_{j}\approx\sqrt{3}\Delta$ and $\omega_{\text{rfd}}\pm\nu_{j}\sim|\gamma_{j}B_{z}|$ 
because of Eq.~(\ref{eq:RWAStrongBzRF}), while the function $F(t)$ carries the frequencies 
$k\omega_{\text{DD}}\sim |\gamma_{j}B_{z}|k/k_{\text{DD}}$. Thus, for $k=k_{\text{DD}}$, we 
need Eq.~(\ref{eq:RWAStrongRFAj}) to eliminate the relevant time-dependent terms, while to 
eliminate the terms having $k\neq k_{\text{DD}}$, the condition (\ref{eq:RWAStrongBzAj}) is 
required. In the case of $k_{\text{DD}}\sim 1$, the condition (\ref{eq:RWAStrongBzAj}) 
becomes Eq.~(\ref{eq:RWAStrongField}). For the addressed spin with $\delta_{n}=0$, we obtain
\begin{equation}
    |\boldsymbol{a}_{n}^{(\pm)}|=\sqrt{\frac{2}{3}(2\mp\sqrt{3})}|\boldsymbol{A}_{n}^{x}|.\label{eq:Abs:apm}
\end{equation}

For the resonance $k_{\text{DD}}\omega_{\text{DD}}=\nu_{n}$  we have the effective Hamiltonian
\begin{equation}
    H_{\text{int}}(t) \approx \frac{m_{s}}{4}f_{k_{\text{DD}}}\sigma_{z}(\boldsymbol{A}_{n}^{z} -\boldsymbol{A}_{n}^{z}\cdot\hat{\nu}_{n}\hat{\nu}_{n})\cdot\boldsymbol{I}_{n},\label{eq:Hint_RfAz}
\end{equation}
when the oscillating terms can be neglected. The Hamiltonians Eq.~(\ref{eq:Hint_Ax}) and
Eq.~(\ref{eq:Hint_RfAz}) allow for the selective detection of individual nuclear spins in the
ensemble. In the next section we will show that by means of an additional rf-field it also
becomes possible to position these nuclear spins.

\subsubsection{Individual control of nuclear spins}
We can selectively control nuclear spins by rf control to allow for the positioning
of the nuclear spin. To this end we apply the rf  control field $H_{\text{rfc}}^{\prime}(t) = \sum_{j}\gamma_{j}V_{\text{rfc}}\cos(\omega_{\text{rfc}}t-\phi_{\text{rfc}})\hat{n}_{\text{rf}}
\cdot\boldsymbol{I}_{j}.$ Similar to the derivation of Eqs.~(\ref{eq:AjTilde}) to (\ref{eq:Az}),
in the rotating frame of $H_{\text{n}}(t)$ the control field
becomes
\begin{equation}
    H_{\text{rfc}}(t) = \sum_{j}\gamma_{j}V_{\text{rfc}}
    \cos(\omega_{\text{rfc}}t-\phi_{\text{rfc}})\tilde{\boldsymbol{n}}_{j}(t)\cdot\boldsymbol{I}_{j},\label{eq:HrfContrl}
\end{equation}
where
\begin{eqnarray}
    \tilde{\boldsymbol{n}}_{j}(t) &=&\tilde{\boldsymbol{n}}_{j}^{x}\cos(\omega_{\text{rfd}}t) +\tilde{\boldsymbol{n}}_{j}^{y}\sin(\omega_{\text{rfd}}t)+\tilde{\boldsymbol{n}}_{j}^{z}(t),\\
    \tilde{\boldsymbol{n}}_{j}^{\alpha} & = & (\boldsymbol{n}_{j}^{\alpha}-\boldsymbol{n}_{j}^{\alpha}\cdot\hat{\nu}_{j}\hat{\nu}_{j})\cos(\nu_{j}t)\nonumber \\
    &  & -\hat{\nu}_{j}\times\boldsymbol{n}_{j}^{\alpha}\sin(\nu_{j}t)+\boldsymbol{n}_{j}^{\alpha}\cdot\hat{\nu}_{j}\hat{\nu}_{j}.
\end{eqnarray}
As in the case for  obtaining $\tilde{\boldsymbol{A}}_{j}(t)$ from $\boldsymbol{A}_{j}$, the vector
$\tilde{\boldsymbol{n}}_{j}(t)$ is generated by rotating  $\hat{n}_{\text{rf}}$ about the axis $\hat{\omega}_{j}$
by an angle $\omega_{\text{rfd}}t$ followed by the second rotation around the axis $\hat{\nu}_{j}$ by an angle $\nu_{j}t$.

When the decoupling control $H_{\text{rfd}}(t)=0$, $\tilde{\boldsymbol{n}}_{j}(t)=\boldsymbol{n}_{j}^{x}\cos(\omega_{j}t)+\boldsymbol{n}_{j}^{y}\sin(\omega_{j}t)+\boldsymbol{n}_{j}^{z}$.
If we choose $\omega_{\text{rfc}}=\omega_{n}$, the rf control field
\begin{equation}
H_{\text{rfc}}(t)\approx\frac{1}{2}\gamma_{n}V_{\text{rfc}}\boldsymbol{n}_{n}(\phi_{\text{rfc}})\cdot\boldsymbol{I}_{n},\label{eq:HrfCtrSimplified}
\end{equation}
assuming that the rf driving is not too strong, i.e., $\gamma_{n}V_{\text{rfc}}\ll\omega_{n}$
and $\gamma_{n}V_{\text{rfc}}\ll\omega_{n}-\omega_{j}$ [see Eq.~(\ref{eq:nj_Phi}) for the expression of $\boldsymbol{n}_{n}(\phi_{\text{rfc}})$].

When the decoupling control $H_{\text{rfd}}(t)\neq0$, $\tilde{\boldsymbol{n}}_{j}(t)$
has components oscillating at the frequency $\nu_{j}$ or $\omega_{\text{rfd}}\pm\nu_{j}$.
We choose weak rf driving $\gamma_{n}V_{\text{rfc}}\ll\sqrt{3}\Delta\ll\omega_{\text{rfd}}$
and $\gamma_{n}V_{\text{rfc}}\ll\omega_{n}-\omega_{j}$. When the rf
frequency is tuned to $\omega_{\text{rfc}}=\omega_{\text{rfd}}\pm\nu_{n}=\omega_{n}+(1\pm\sqrt{3})\Delta$,
we have the effective single spin control
\begin{equation}
    H_{\text{rfc}}(t)\approx\frac{\gamma_{n}V_{\text{rfc}}}{4}\boldsymbol{b}_{n}^{(\pm)}\cdot\boldsymbol{I}_{n},
\label{eq:HrfCtrSimplifiedRF}
\end{equation}
\begin{eqnarray}
    \boldsymbol{b}_{n}^{(\pm)} & \equiv & \boldsymbol{n}_{n}(\phi_{\text{rf}}) -\boldsymbol{n}_{n}(\phi_{\text{rf}})\cdot\hat{\nu}_{n}\hat{\nu}_{n}\pm\hat{\nu}_{n}\times
    \boldsymbol{n}_{n}(\phi_{\text{rf}}+\frac{\pi}{2}).\label{eq:VecOmegaPM}
\end{eqnarray}
While choosing $\omega_{\text{rfc}}=\nu_{j}$, we obtain
\begin{equation}
    H_{\text{rfc}}(t)\approx\frac{1}{2}\gamma_{n}V_{\text{rfc}}\boldsymbol{b}_{n}^{(z)}\cdot\boldsymbol{I}_{n},
\end{equation}
\begin{equation}
    \boldsymbol{b}_{n}^{(z)} =(\boldsymbol{n}_{n}^{z}-\boldsymbol{n}_{n}^{z}\cdot\hat{\nu}_{n}\hat{\nu}_{n}) \cos\phi_{\text{rfc}}-\hat{\nu}_{n}\times\boldsymbol{n}_{n}^{z}\sin\phi_{\text{rfc}}.
\end{equation}
In the next section we show how with these abilities of selectivity and tunable control on individual nuclear
spins we can measure the positions of individual nuclei with high precision.

\section{The protocol at work: Measurement of Hyperfine Interactions and Nuclear Positioning}\label{sec:Positioning}

Our protocol measures the hyperfine vectors $\boldsymbol{A}_{j}$ and derives
the positions of nuclei by studying the response of the electron spin quantum coherence
to changes of the available control parameters. In this section we detail the individual
steps of the protocol and demonstrate its effectiveness by means of numerical examples for
realistic parameter sets.

We initialize the NV electron spin in a coherent superposition state $|\psi_{x}\rangle\equiv \frac{1}{\sqrt{2}}\left(|m_{s}\rangle+|0\rangle\right)$ with $m_{s}=\pm1$ by optical initialization
followed by a $\pi/2$ pulse \cite{doherty2013nitrogen,dobrovitski2013quantum}. For temperatures
significantly exceeding the nuclear spin Zeeman energy ($<\text{MHz\ensuremath{\sim\mu}K}$), the
nuclear spin state is well approximated by $\rho_{B}\propto I_{B}$ where $I_{B}$ is the identity
operator. Hence the initial state is $\rho(0) \propto |\psi_{x}\rangle\langle\psi_{x}|\otimes I_B$.
Following this initialization procedure, we apply DD sequences and the rf fields. The population
in the state $|\psi_{x}\rangle$, given by $p_{|\psi_{x}\rangle}(t) = \text{Tr}\left[\rho(t) |\psi_x\rangle\langle\psi_x|\right]$,
changes when the NV electron spin interacts with the nuclei. We use $L_{0,m_{s}}(t)=2p_{|\psi_{x}\rangle}(t)-1$
to describe the signal. In terms of the states $|m_{s}\rangle$ and $|0\rangle$, the quantity $L_{0,m_{s}}(t)$
is the off-diagonal part of the reduced NV electron density matrix and represents the quantum coherence~\cite{Maze2008PRB,zhao2012sensing}.

\subsection{3D positioning for non-interacting spins \label{subsec:IndividualSpin}}

\begin{figure}
\includegraphics[width=1\columnwidth]{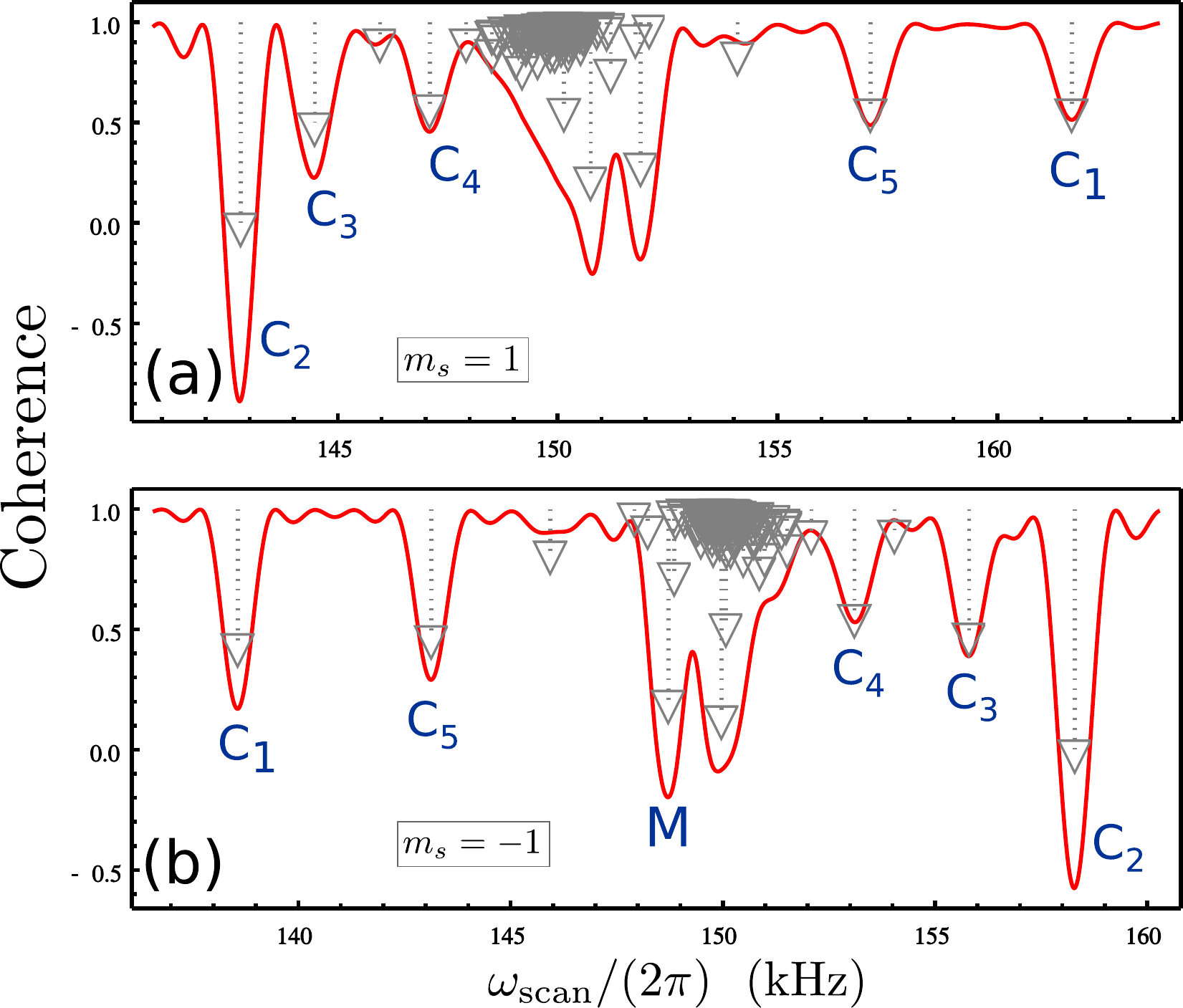}\caption{\label{fig:FigSpectrum}
(color online). Resonance patterns for a diamond sample containing
$442$ $^{13}\text{C}$ nuclei on randomly chosen lattice sites using
the AXY sequences. The static magnetic field $B_{z}=140.1$ G and the
AXY sequences utilize the first harmonic ($k_{\text{DD}}=1$) and $f_{1}=0.06$.
(a) and (b) corresponds to the choice of $m_{s}=+1$ or $m_{s}=-1$
NV state, respectively. The down arrows (with empty triangles as
arrow heads) indicate the locations of $\omega_{j}$, and the lengths
of the arrows measure the relative coupling strengths $|\boldsymbol{A}_{j}^{x}|$.
An addressed spin (the $n$-th spin) reduces the coherence by the factor given by
Eq.~(\ref{eq:Lcoher_Ax}), which appears as coherence dips on the spectra. }
\end{figure}

For spins such that their dipolar coupling $H_{\text{nn}}$ is negligible, we can
individually address them without applying the rf decoupling field in Eq.~(\ref{eq:Hrfd}).
To illustrate this situation we  have performed simulations on a diamond sample containing
442 $^{13}\text{C}$ spins which have been randomly generated on lattice sites of the
diamond structure according to an abundance of 0.5 \%. The simulations are carried
out by disjoint cluster expansion including dipolar interactions up to a group size
of six $^{13}\text{C}$ \cite{Maze2008PRB} and checked with cluster correlation expansion
method \cite{Yang2008PRB}. We demonstrate the measurement of the hyperfine fields
$\boldsymbol{A}_{j}$ and hence the positions of non-interacting nuclear spins.

\subsubsection{Addressing a single spin for measurement}
To measure the hyperfine vectors $\boldsymbol{A}_{j}$, we individually address the nuclear
spins. According to the theory presented in Sec. \ref{sub:addressing}, we tune the frequency
$k_{\text{DD}}\omega_{\text{DD}}=\omega_{\text{scan}}$ by changing the DD pulse intervals.
When $\omega_{\text{scan}}$ matches the resonance frequency of single nuclei, i.e.,
$\omega_{\text{scan}}=\omega_{j}$, the coupling Hamiltonian in Eq.~(\ref{eq:Hint_Ax})
is achieved. This interaction entangles the NV electron and the addressed nuclei which
induces a dip on the NV coherence $L_{0,m_{s}}$. When only the $n$-th spin is addressed,
from Eq.~(\ref{eq:Hint_Ax}), the coherence reads
\begin{equation}
    L_{0,m_{s}}(t)=\cos\left(\frac{1}{4} f_{k_{\text{DD}}} |\boldsymbol{A}_{n}^{x}| t \right). \label{eq:Lcoher_Ax}
\end{equation}
In Fig.~\ref{fig:FigSpectrum}, we apply a magnetic field $B_{z}=140.1$ G (i.e.,
$\gamma_{j}B_{z}=2\pi\times150$ kHz for $^{13}\text{C}$ spins) along the NV axis,
which is much stronger than the typical hyperfine fields $|\boldsymbol{A}_{j}|$ in
the sample. To fulfill the conditions of Eqs.~(\ref{eq:RWAStrongField}) and
(\ref{eq:RWAWeekFdd}), we have used $k_{\text{DD}}=1$ and $f_{1}=0.02$ in the
expansion, see Eq.~(\ref{eq:Ft-Expansion}). Here the AXY sequences have $f_{k}=0$,
$k$ being $3$ or any even number. To achieve a total evolution time of around
1 ms, $1520$ pulses are used in the AXY sequences. (Note that, in total $304$ robust
composite X and Y pulses are applied given that each composite pulse consists of 5
elementary $\pi$ pulses in the AXY sequences. Each DD period $2\pi/\omega_{\text{DD}}$
contains two composite pulses while a minimal control circle in an AXY-8 sequence has
8 composite pulses.) By scanning the DD frequency $\omega_{\text{DD}}$ for
different values, clear coherence dips, which are signatures of spin addressing,
appear, as shown in Fig.~\ref{fig:FigSpectrum}. These dips locate at the Larmor
frequencies $\omega_j$ of the nuclear spins, as indicated by the down arrows on
the figures. Each of the coherence dips marked by $\text{C}_{j}$ on the figures
is caused by coupling to a single nuclear spin, which can be confirmed by the
sinusoidal coherence oscillations when changing the total evolution time \cite{taminiau2012detection,loretz2015spurious}.
However, the total DD sequence time has to be integer multiples of the DD pulse interval,
which is synchronized with the resonant frequencies $\omega_{j}=k_{\text{DD}}\omega_{\text{DD}}$.
Therefore, for a fixed magnetic field $B_{z}$, continuous plot of the coherence as a function
of the total sequence time is impossible.
\begin{figure}
\includegraphics[width=0.95\columnwidth]{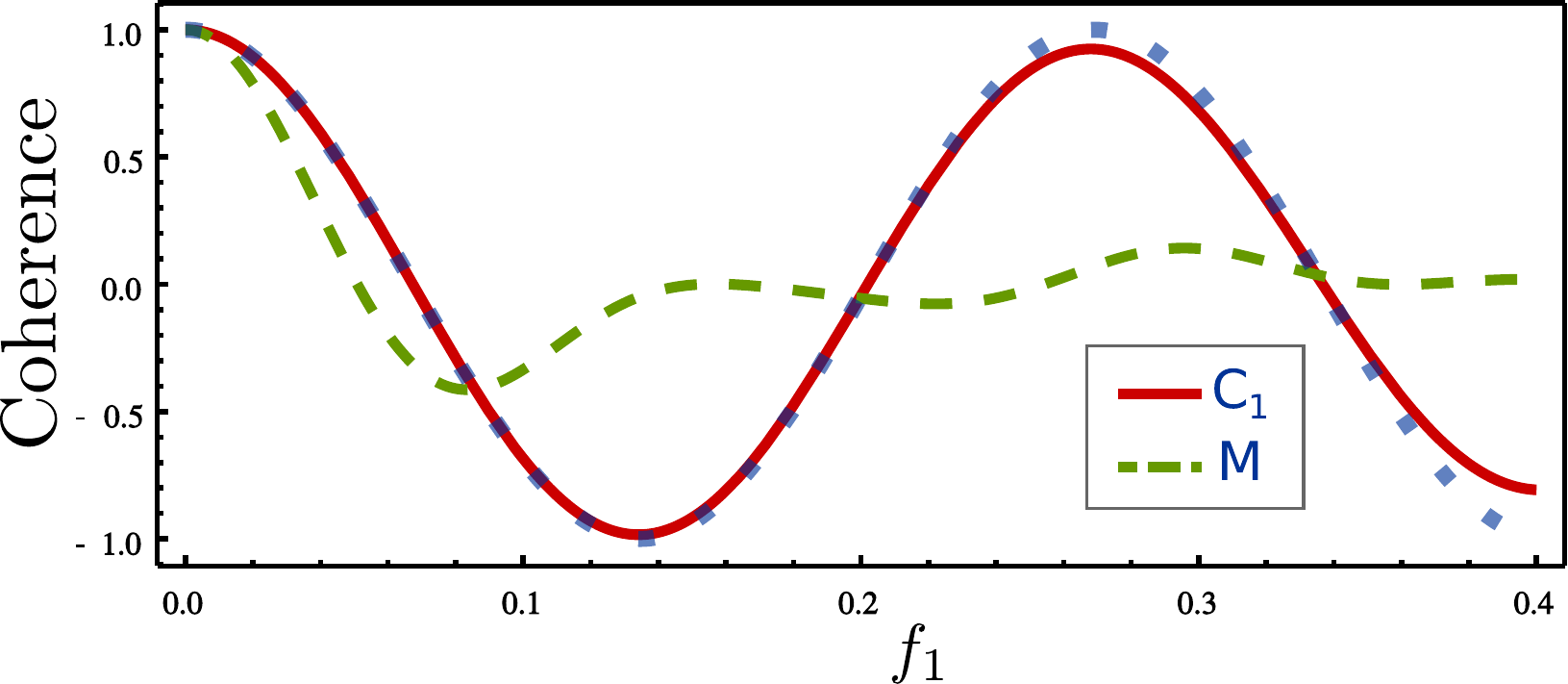}\caption{\label{fig:Figf1}
(color online). Coherence oscillations as a function of $f_{1}$. Here $f_{1}$
by the AXY sequences can be tuned continuously and other parameters are the same as
those in Fig.~\ref{fig:FigSpectrum}~(b). The red solid line shows the sinusoidal
coherence oscillation when $\omega_{\text{scan}}$ is set to the resonant frequency
of the addressed spin denoted by $\text{C}_{1}$ in Fig.~\ref{fig:FigSpectrum}~(b),
and the blue dots mark the curves calculated by Eq.~(\ref{eq:Lcoher_Ax}). Green dashed
line is the case that $\omega_{\text{scan}}$ is set to the frequency dip denoted by
M in Fig.~\ref{fig:FigSpectrum}~(b), where a few spins contribute the signal.}
\end{figure}
We propose another way to confirm that the coherence dips $\text{C}_{j}$ are caused by coupling to a single nuclear
spin. With the ability of modifying the $f_{1}$ coefficient by the AXY sequences,
we can tune the NV-nucleus coupling strength continuously and linearly with $f_{1}$,
and therefore obtain continuous change of the coherence as a function of $f_{1}$
(instead of total sequence time).
The red solid line in Fig.~\ref{fig:Figf1} shows the coherence when we continuously change the value of $f_{1}$ while keeping other parameters unchanged and setting the frequency $\omega_{\text{scan}}$ to the resonant frequency of a single nuclear spin. We can see clear sinusoidal coherence oscillation which fits the
dynamics of a single spin. Note that there is a slight decay of the oscillation with increasing
$f_{1}$, because a relatively large $f_{1}$ will violate the conditions
Eqs.~(\ref{eq:RWAStrongField}) and (\ref{eq:RWAWeekFdd}). This oscillation
decay can be reduced by using a longer evolution time. For the coherence
dip denoted by M in Fig.~\ref{fig:FigSpectrum}~(b), changing
$f_{1}$ do not produce sinusoidal coherence oscillation, as shown by the green dashed line in
Fig.~\ref{fig:Figf1}, since the NV electron spin couples to more
than one nuclear spins at the position M, at which we can see several
resonant frequencies $\omega_{j}$ indicated in Fig.~\ref{fig:FigSpectrum}~(b).
Without the ability of tuning $f_{k}$, e.g., by using CPMG or XY
sequences, the coupling between NV electron and nuclear spins can not
be tuned, and this limits the fidelity of two-qubit quantum gates
since the interaction time has to be integer multiples of $\pi/\omega_{j}$,
which is fixed for the nuclear spins at a magnetic field. The oscillations
of the coherence $L_{0,m_{s}}$ tuned by $f_{1}$ demonstrate that we can finely change
the interaction strength and hence enable high-fidelity two-qubit
quantum gates between the NV electron and nuclei. For example, for
the point with vanishing coherence $L_{0,m_{s}}$ in Fig.~\ref{fig:Figf1}~(a), the
NV electron spin and the addressed spin are maximally entangled.

\subsubsection{Strength of hyperfine field}
After identifying the resonance frequencies $\omega_{j}$ by individual spin
addressing, we can obtain the strengths of the hyperfine fields. The Larmor
frequency $\omega_{j} = \sqrt{(\gamma_{j}B_{z}-\frac{m_{s}}{2}A_{j}^{\parallel})^{2} +
\frac{1}{4}(A_{j}^{\perp})^{2}}$ is a function of the parallel component
$A_{j}^{\parallel}=\boldsymbol{A}_{j}\cdot\hat{z}$ of the hyperfine field
along the $\hat{z}$ direction and the strength of perpendicular part
$A_{j}^{\perp}=|\boldsymbol{A}_{j}-A_{j}^{\parallel}\hat{z}|$. Under a strong
magnetic field we have $\omega_{j}\approx|\gamma_{j}B_{z}-\frac{m_{s}}{2}A_{j}^{\parallel}|$.
In this way the coherence dips are approximately symmetric about $\gamma_{j}B_{z}$
under the transformation $m_s \rightarrow -m_s$ (although not exactly because
of the direction and strength of $\boldsymbol{\omega}_{j}$ depend on $m_{s}$)
see  Figs. \ref{fig:FigSpectrum}~(a) $m_{s}=1$ and (b) $m_{s}=-1$. By measurements
of $\omega_{j}$ using different NV states $|m_{s}\rangle$ or under different
magnetic fields $B_{z}$, we obtain $A_{j}^{\parallel}$ and $A_{j}^{\perp}$
using the expression of $\omega_{j}$. Note that we can also measure $|\boldsymbol{A}_{n}^{x}|$
by fitting the coherence data in Fig.~\ref{fig:Figf1} with Eq.~(\ref{eq:Lcoher_Ax}).
Under a strong magnetic field $|\boldsymbol{A}_{n}^{x}|\approx A_{n}^{\perp}$.

\begin{figure}
\includegraphics[width=0.95\columnwidth]{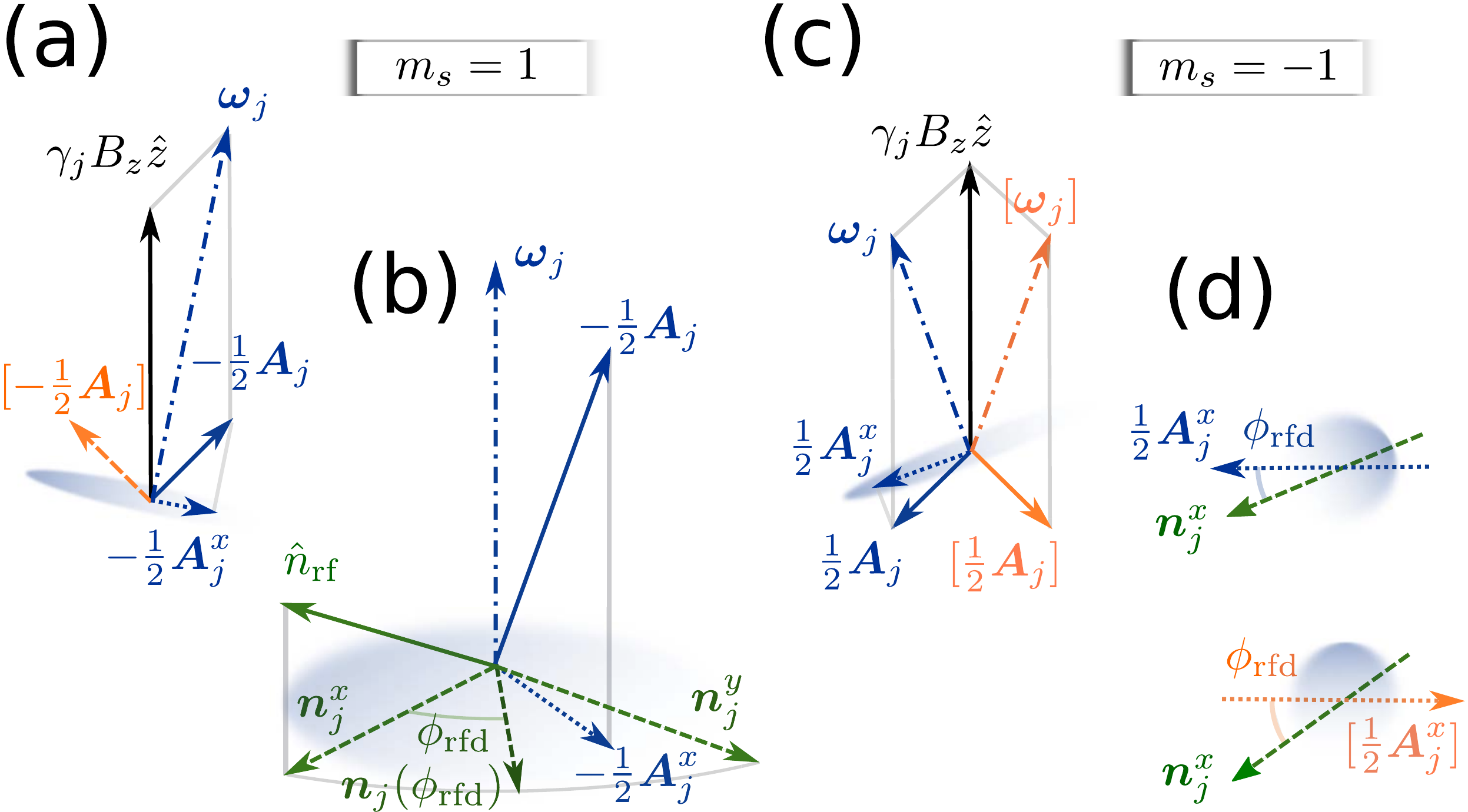}\caption{\label{fig:FigSketch}
(color online). Illustration of measuring the direction of $\boldsymbol{A}_{j}$ by
using a rf control field to break the system symmetry. (a) and (b) sketch the vector
relations with $m_{s}=1$, while for (c) and (d) with $m_{s}=-1$. (a) and (b): the
first scan of $\phi_{\text{rfc}}$ (e.g., with using $m_{s}=1$) left two possible
directions of $\boldsymbol{A}_{j}$ using Eq.~(\ref{eq:parallel}) {[}see (a) with
the non-physical one denoted with ``{[}{]}'' in orange{]}, because of the unknown
sign in $\pm\boldsymbol{A}_{j}^{x}$ {[}see (b){]}. (c) and (d): using another scan
with different control parameters (e.g., changing $m_{s}$ to $m_{s}=-1$), the physical
solution of $\boldsymbol{A}_{j}$ is picked up. (c) shows the two possible $\omega_{j}$
for the two possible directions of $\boldsymbol{A}_{j}$ obtained from the first scan
of $\phi_{\text{rfc}}$. (d) shows that the angles between $\boldsymbol{n}_{j}^{x}$ and
$\boldsymbol{A}_{j}^{x}$ are different when lying on the corresponding plane that is
perpendicular to $\boldsymbol{\omega}_{j}$ or the non-physical one ${[}\boldsymbol{\omega}_{j}{]}$.
Therefore, from the parallel condition Eq.~(\ref{eq:parallel}) in the second scan
with a different $m_{s}$ from the one in the first scan, the actual direction of
$\boldsymbol{A}_{j}^{x}$ is obtained. With the measured values of $A_{j}^{\parallel}$
and $A_{j}^{\perp}$, full information of $\boldsymbol{A}_{j}$ is measured. }
\end{figure}

\subsubsection{Direction of hyperfine field}
The direction of $\boldsymbol{A}_{j}$ (note that $\boldsymbol{A}_{j}$ is a vectorial quantity)
projected on the plane perpendicular to the magnetic field $B_{z}\hat{z}$ can not be inferred
by only reconstructing $\omega_{j}$, i.e. by measuring  $A_{j}^{\parallel}$ and $A_{j}^{\perp}$.
In this respect a method to estimate the relative directions of $\boldsymbol{A}_{j}$ has been
recently proposed  by utilizing nucleus-nucleus interactions \cite{ajoy2015atomic} which, however,
could be inaccurate because of weak dipolar coupling between the nuclei and the complicated
dynamics of many-spin clusters.

\begin{figure}
\includegraphics[width=0.95\columnwidth]{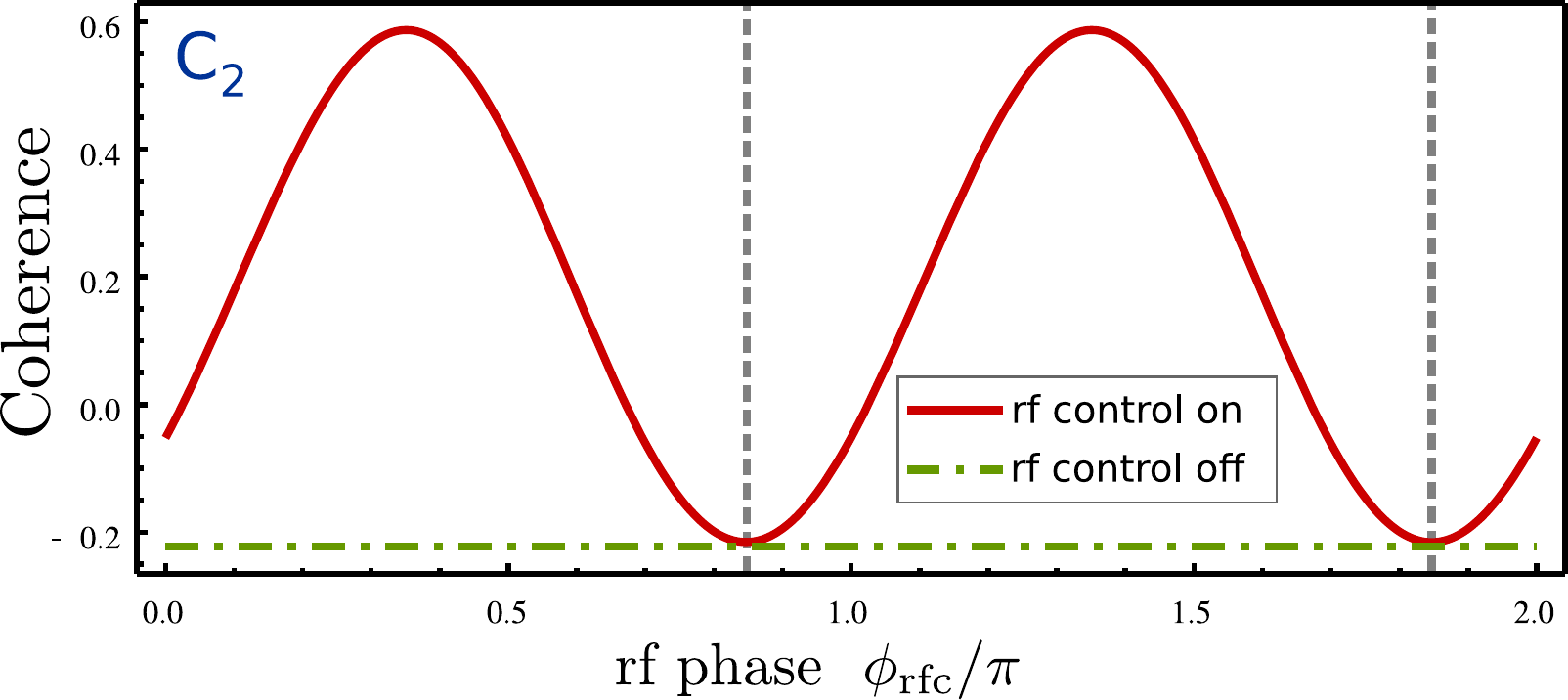}\caption{\label{fig:FigAngle}
(color online). Coherence $L_{0,m_{s}}$ as a function of rf control phase
$\phi_{\text{rfc}}$. We set $\omega_{\text{scan}}$ at the resonant frequency denoted
by $\text{C}_{2}$ in Fig.~\ref{fig:FigSpectrum}~(a) and use $f_{1}=0.04$ for the
AXY sequences. Other parameters are the same as those used in Fig.~\ref{fig:FigSpectrum}~(a).
Red line is the coherence that a rf control field is applied with a driving amplitude
$\gamma_{j}V_{\text{rfc}}=2\pi\times1$ kHz for the $^{13}\text{C}$
spins. Green dashed dotted line is the coherence without rf control field
($V_{\text{rfc}}=0$). The vertical dashed lines denote the cases that $\boldsymbol{n}_{j}(\phi_{\text{rfc}})$
and $\boldsymbol{A}_{j}^{x}$ are parallel.}
\end{figure}

Here, we show that the direction of
$\boldsymbol{A}_{j}$ can be measured by simply using the rf control field
in Eq.~(\ref{eq:HrfContrl}). Note that both the effective field direction
$\boldsymbol{n}_{j}(\phi_{\text{rfc}})$ [see Eqs.~(\ref{eq:HrfCtrSimplified}) and (\ref{eq:nj_Phi})] and $\boldsymbol{A}_{j}^{x}$
in Eq.~(\ref{eq:Hint_Ax}) lie in a plane perpendicular to $\hat{\omega}_{j}$,
see Figs. \ref{fig:FigSketch}~(a) and (b). The orientation of the vector $\boldsymbol{n}_{j}(\phi_{\text{rfc}})$
in the plane orthogonal to $\hat{\omega}_{j}$ can be controlled by the phase $\phi_{\text{rfc}}$.
When $\boldsymbol{n}_{j}(\phi_{\text{rfc}})$
and $\boldsymbol{A}_{j}^{x}$ are parallel, i.e.,
\begin{equation}
\boldsymbol{n}_{j}(\phi_{\text{rfc}})\times\boldsymbol{A}_{j}^{x}=0,\label{eq:parallel}
\end{equation}
the rf control Hamiltonian $H_{\text{rfc}}(t)$ {[}see Eq.~(\ref{eq:HrfCtrSimplified}){]}
commutes with the interaction Hamiltonian $H_{\text{int}}$ {[}see Eqs.~(\ref{eq:Hint_Ax})
and (\ref{eq:HrfCtrSimplified}) for the case without a rf decoupling field and Eqs.~(\ref{eq:Hint_RfApm})
and (\ref{eq:HrfCtrSimplifiedRF}) when a rf decoupling field has been applied{]} and application
of $H_{\text{rfc}}(t)$ has no effect on the coherence of NV electron
spin.
Eq.~(\ref{eq:parallel}) involves the values of the vectors
$\hat{n}_{\text{rf}}$ and $\boldsymbol{A}_{j}$, as well as the control
parameters magnetic field strength $B_{z}$ and magnetic number $m_{s}$.
When $\boldsymbol{n}_{j}(\phi_{\text{rfc}})$ and $\boldsymbol{A}_{j}^{x}$
are not parallel, the rf control field reduces the interaction between
the NV electron and the addressed nuclear spin. This reduction of
interaction is maximum when $\boldsymbol{n}_{j}(\phi_{\text{rfc}})$
and $\boldsymbol{A}_{j}^{x}$ are perpendicular. Therefore, we can
measure the solutions $\phi_{\text{rfc}}$ to Eq.~(\ref{eq:parallel}),
by which we can infer the direction of $\boldsymbol{A}_{j}$ relative
to the NV axis $\hat{z}$ and $\hat{n}_{\text{rf}}$. An example of
the coherence for a certain final evolution time as a function of phase $\phi_{\text{rf}}$ is shown
in Fig.~\ref{fig:FigAngle}, where vertical lines indicates the situation
that $\boldsymbol{n}_{j}(\phi_{\text{rfc}})$ and $\boldsymbol{A}_{j}^{x}$
are parallel. Eq.~(\ref{eq:parallel}) is invariant under the transformation
$\phi_{\text{rfc}}\rightarrow\phi_{\text{rfc}}+\pi$ (i.e., $\hat{n}_{\text{rf}}\rightarrow-\hat{n}_{\text{rf}}$),
which implies that a single scan of $\phi_{\text{rfc}}$ left two possible
results $\pm\boldsymbol{A}_{j}^{x}$. Additionally we want to comment that the role of the rf field in our protocol is significantly different to the one in  \cite{laraoui2015imaging} where a rf $\pi/2$ pulse is used to complete the polarization transfer between the NV center and a target spin.

With two possible $\pm\boldsymbol{A}_{j}^{x}$, we obtain
two possible hyperfine fields  $\boldsymbol{A}_{j}$, see Fig.~\ref{fig:FigSketch}~(a) for the actual $\boldsymbol{A}_{j}$ and its non-physical mirror solution.
The sign ambiguity of $\pm\boldsymbol{A}_{j}^{x}$
can be solved by noting that $\boldsymbol{A}_{j}$ and its non-physical mirror solution generally have different relative angles with respect to $\hat{n}_{\text{rf}}$.
For alignments that $\hat{n}_{\text{rf}}$
is not perpendicular to the plane spanned by $\hat{z}$ and $\boldsymbol{A}_{j}^{x}$,
as shown in Fig.~\ref{fig:FigSketch}, the angle between the projected vector $\boldsymbol{n}_{j}(0)$ and
$\boldsymbol{A}_{j}^{x}$ {[}i.e., the solution $\phi_{\text{rfc}}$
to Eq.~(\ref{eq:parallel}){]} changes when changing $\hat{\omega}_{j}$,
which depends on the values of magnetic number $m_{s}$ and magnetic
field strength $B_{z}$. 

\subsubsection{Measured positions}

Therefore, we can distinguish the signs of $\pm\boldsymbol{A}_{j}^{x}$ (hence the actual $\boldsymbol{A}_{j}$ and its non-physical mirror solution)
by checking Eq.~(\ref{eq:parallel}) for other
parameters $m_{s}$ and $B_{z}$.
Here the changes of $m_{s}$ already provide enough information to select the right
sign of $\pm\boldsymbol{A}_{j}^{x}$. With the measured values of
$A_{j}^{\parallel}$, $A_{j}^{\perp}$, and the direction of $\boldsymbol{A}_{j}^{x}$,
the hyperfine field $\boldsymbol{A}_{j}$ is fully determined.
Since the scheme does not rely on the amplitudes of the rf field $V_{\text{rfc}}$,
amplitude uncertainties do not affect the measurements.

\begin{figure}
\includegraphics[width=1\columnwidth]{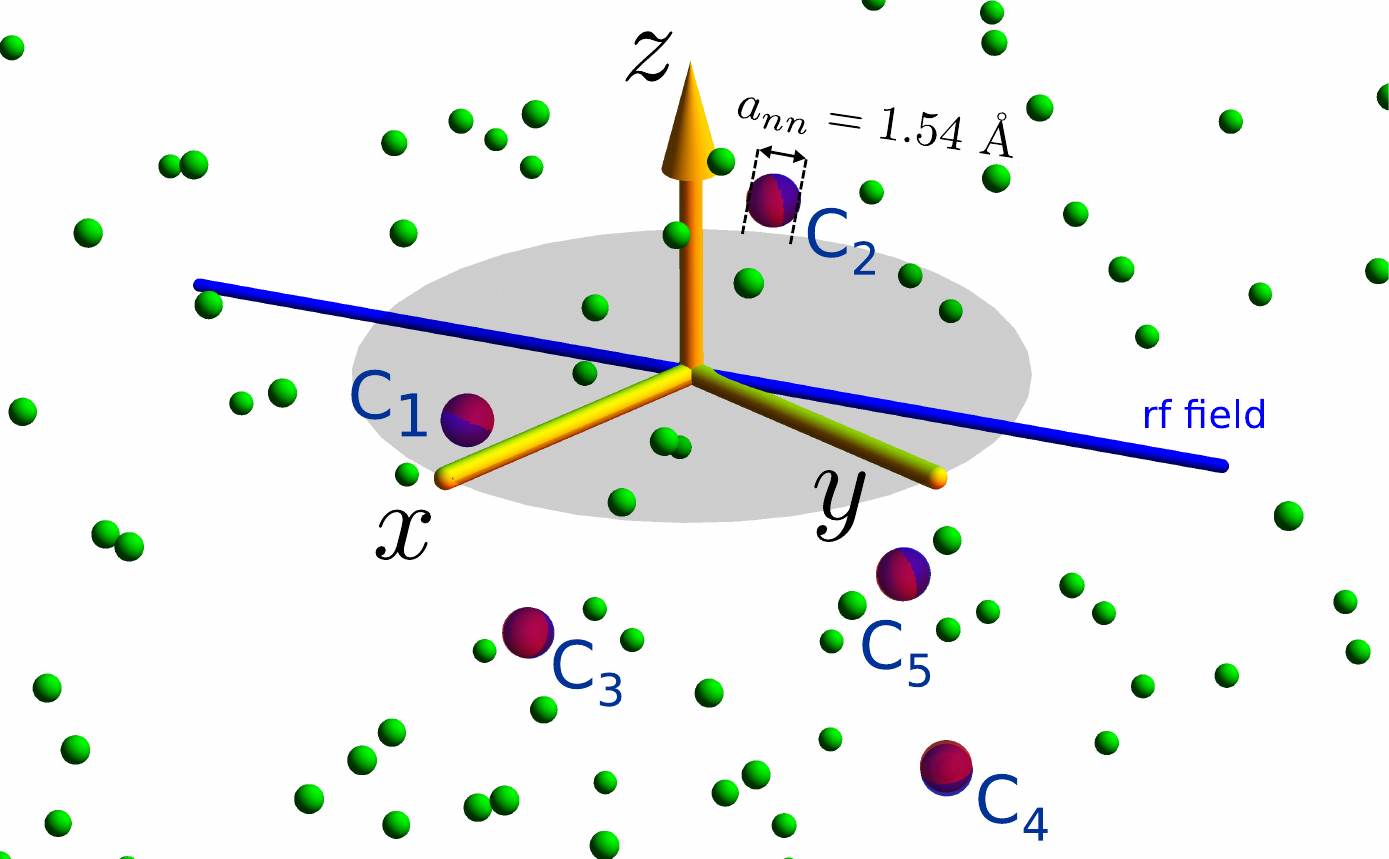}\caption{\label{fig:Fig3D}
(color online). Comparison of the measured (red
spheres) and actual (blue spheres) $^{13}\text{C}$ positions, indicated
by spheres of the diameter $a_{nn}=1.54\:\text{\ensuremath{\AA}}$
that is the nearest neighbour distance of carbon atoms on the diamond lattice.
Measurement uncertainties smaller than $0.5 a_{nn}$ can already
fix the positions of $^{13}\text{C}$ in the lattice sites of diamond.
Their measured distances $r_{j}$ rang from 0.92 to 1.38 nm. Other green
spheres of a smaller size are other $^{13}\text{C}$ spins forming
a spin bath. The shaded disk has a radius of 1 nm. The measured spin
denoted by $\text{C}_{1}$ defines the coordinate which makes the
$\text{C}_{1}$ spin lie in the $x-z$ plane. The blue line indicates
the measured rf field direction $\pm\hat{n}_{\text{rf}}$, which is
almost parallel (indistinguishable) to the actual one. }
\end{figure}

The scheme measures the relative directions among $\hat{z}$, $\hat{n}_{\text{rf}}$, and
the  hyperfine fields $\boldsymbol{A}_{j}$. If the direction
of rf field $\hat{n}_{\text{rf}}$ is unknown, we can define
a coordinate system $\mathbb{S}$ by $\hat{z}$ and
the hyperfine field $\boldsymbol{A}_{j}$ of one nuclear spin.
In terms of $\mathbb{S}$, $\hat{n}_{\text{rf}}$ is fixed and
the hyperfine fields of other spins are measured in the coordinate $\mathbb{S}$.
In our simulation, we assume unknown $\hat{n}_{\text{rf}}$ and define
the coordinate $\mathbb{S}$ that the hyperfine field $\boldsymbol{A}_{j}$ of the nucleus $\text{C}_{1}$
in Fig.~\ref{fig:FigSpectrum} is aligned in the $x-z$ plane with $\boldsymbol{A}_{j}\cdot \hat{x}>0$ (see Fig.~\ref{fig:Fig3D} for the coordinate).

For nuclear spins not too close
to the NV center (e.g., the case in our simulations), the hyperfine interaction  $H_{\text{hf}}$ takes the dipolar form and
\begin{equation}
\boldsymbol{A}_{j}=\frac{\mu_{0}}{4\pi}\frac{\gamma_{e}\gamma_{j}}{r_{j}^{3}}\left(\hat{z}-\frac{3\hat{z}\cdot\boldsymbol{r}_{j}\boldsymbol{r}_{j}}{|\boldsymbol{r}_{j}|^{2}}\right).\label{eq:Aj}
\end{equation}
From the position dependence of $\boldsymbol{A}_{j}$, we obtain the nuclear positions
$\boldsymbol{r}_{j}$ by solving Eq.~(\ref{eq:Aj}).
Note that the hyperfine interaction $H_{\text{hf}}$ is symmetric
under placing the nuclei at the opposite directions $\boldsymbol{r}_{j}\rightarrow-\boldsymbol{r}_{j}$.
Therefore determining $H_{\text{hf}}$ alone is not sufficient to
distinguish the ambiguity of the sign in $\pm\boldsymbol{r}_{j}$.
Using homogeneous magnetic fields along different directions cannot solve this
ambiguity, since the Zeeman term $\gamma_{j}\boldsymbol{B}\cdot \boldsymbol{I}_{j}$
is invariant under $\boldsymbol{r}_{j}\rightarrow-\boldsymbol{r}_{j}$. However,
we can solve this ambiguity by introducing a gradient magnetic field along the
NV axis $B_{z}=B_{z}(\boldsymbol{r})$ to break the symmetry. When we know that
the detected spins are placed at some regions, e.g., on top of the diamond surface
(hence NV center), we do not need to distinguish $\pm\boldsymbol{r}_{j}$.

From the measured $\boldsymbol{A}_{j}$, we obtain the positions of the nuclear
spins with distances $r_{j}$ ranging from 0.92 to 1.38 nm. In Fig.~\ref{fig:Fig3D},
we show the actual positions (blue spheres) and the measured ones (red spheres)
that are obtained by the measured hyperfine fields.

\subsection{Individual 3D spin positioning for interacting spin clusters} \label{subsec:C13Cluster}
In the previous subsection we have demonstrated the ability of our protocol to deliver
positioning of nuclear spins with Angstrom precision in non-interacting spin clusters.
In this subsection we are now turning to the case of interacting nuclear spin clusters.
The procedure to measure the hyperfine fields $\boldsymbol{A}_{j}$ and therefore the
positions of the nuclear spins is quite similar to that described in the previous subsection~\ref{subsec:IndividualSpin}.

To show the working principles of our method in this case, we consider a
cluster of three coupled nuclear spins (see Table~\ref{table:clusterPositions})
of which the internuclear distances are $a_{nn}$, $a_{nn}$, and $2\sqrt{2/3}a_{nn}$.
Here $a_{nn}=1.54$ $\text{\AA}$ is the nearest neighbor distance between carbon
nuclei in the diamond lattice. With a suitably tuned rf decoupling field we can
suppress the internuclear dipolar coupling (see \cite{cai2013large} and
Sec.~\ref{subsubsec:DecouplingNN}) and therefore enable individual spin addressing and
positioning. We use a rf decoupling field with $\Delta=20$ kHz to suppress internuclear
dipolar coupling. This gives rise to the application of an rf field such that its Rabi frequency is $\approx \sqrt{2} \Delta$ which implies a decoupling field intensity of $\approx 28$ G. Note that control fields with intensities around $0.1$ T have already been implemented~\cite{michal2008two,fuchs2009gigahertz}. In addition, unlike traditional nuclear magnetic resonance in solids that requires switching off of rf control fields to allow for detection windows, a stable rf decoupling field can be turned on during the whole period of our protocol, including NV electron spin initialization and readout, because the frequency of rf decoupling field is far off-resonance to the transition frequencies of the NV electron spin. Therefore, fast switching of rf decoupling is not required in our protocol, and we can use external coils that can avoid possible heating on the diamond sample.
The Larmor frequency $\omega_{j}$ should be much larger than $\Delta$ to
make the rf decoupling control possible. Thus we choose a strong magnetic field with $\gamma_{j}B_{z}=900$ kHz
for $^{13}\text{C}$ spins. To achieve a total evolution time of around $1$ ms, the third
harmonic ($k_{\text{DD}}=3$) and 3040 pulses (i.e., 608 composite pulses) are used for the
AXY sequences with $f_{3}=0.06$.

\begin{table}[]
\centering
\caption{
Nuclear positions in an interacting $^{13}\text{C}$ spin cluster as
well as their measured values.
The direction {[}1,1,1{]} is chosen as the NV symmetry axis
and the NV center is located at {[}0,0,0{]}.
}
\label{table:clusterPositions}
\begin{tabular}{l|ccc|ccc}
    			 & \multicolumn{3}{c|}{Nuclear positions ($\AA$)} & \multicolumn{3}{c}{Measured values ($\AA$)} \\ \hline
$\text{Q}_{1}$   & {[}3.57, & 10.71, & 3.57{]}  &  {[}3.56, & 10.69, & 3.55{]}        \\
$\text{Q}_{2}$   & {[}4.46, & 9.82, & 4.46{]}   &  {[}4.32, & 9.82, & 4.46{]}      \\
$\text{Q}_{3}$   & {[}3.57, & 8.93, & 5.36{]}   &  {[}3.65, & 8.96, & 5.23{]}
\end{tabular}
\end{table}

\begin{figure}
\includegraphics[width=0.95\columnwidth]{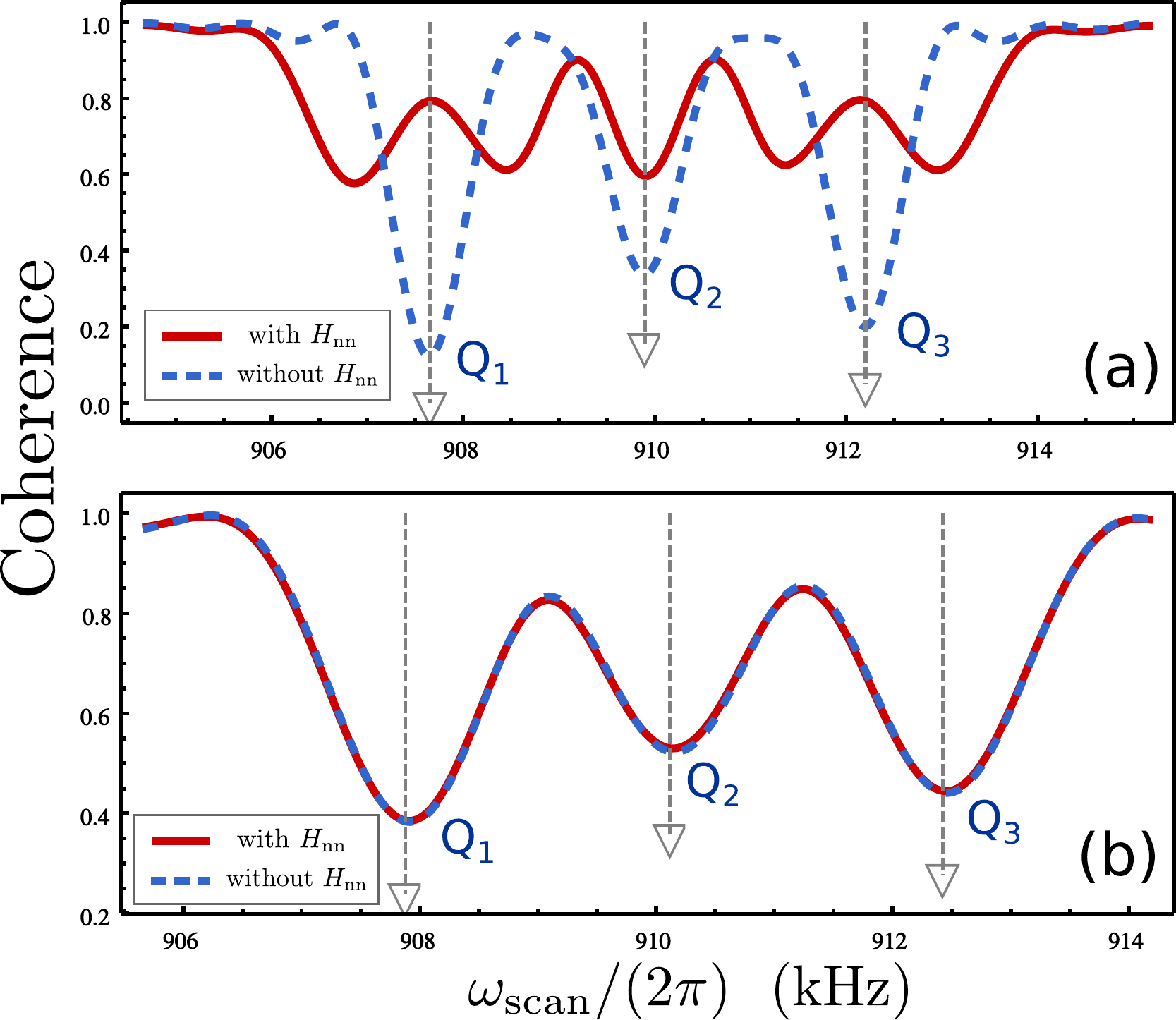}\caption{\label{fig:FigSpectrumMg}(color online).
Resonance patterns for a
coupled cluster consisting of three $^{13}\text{C}$ spins in diamond shown in Table~\ref{table:clusterPositions}).
The magnetic field $B_{z}\approx 840.7$ G
is applied along the NV symmetry axis and we choose $m_{s}=-1$.
The vertical down arrows indicate the locations of single spin
resonant frequencies, while their lengths denote the relative coupling
strengths. The red solid line in (a) is the case without rf decoupling
field, while the red line in (b) is the case that a rf decoupling
field with $\Delta=2\pi\times20$ kHz is applied to suppress nuclear
dipolar coupling. The blue dashed lines in (a) and
(b) are the corresponding simulations without considering nuclear
dipolar interaction. The application of rf decoupling reduces the electron-nuclear interactions by a factor $1/\sqrt{3}$ and hence reduces the separation between resonances.}
\end{figure}

In Fig.~\ref{fig:FigSpectrumMg}~(a), we plot the coherence
(red line) using the method that we have described in the previous subsection,
i.e., without the application of rf decoupling field, which deviates considerably
from the blue dashed line which is obtained by neglecting the internuclear interactions
$H_{\text{nn}}$. Indeed, a comparison with the blue dashed line shows that
from the red coherence profile positioning of the nuclear spins in the spin
cluster is not possible without the assistance of further numerical efforts invested to decode the spectrum. 
Although computationally feasible for the present case, such a numerical decoding rapidly becomes computationally challenging for even a 
moderate number of spins (see for example the case of malic acid studied in Section \ref{malate}) especially when compared with the 
analysis of the signal provided by our method that can be effectively considered as generated from a set of non-interacting nuclear spins [see Fig.~\ref{fig:FigSpectrumMg}~(b)].

To demonstrate individual addressing and positioning of nuclear spins while
suppressing $H_{\text{nn}}$, we choose the control frequencies Eqs.~(\ref{eq:ResRfd})
and (\ref{eq:ResWdd}). When $\omega_{\text{scan}}=\omega_{j}$, we address the $j$-th
nuclear spin with the interaction Hamiltonian Eq.~(\ref{eq:Hint_RfApm}). With the
help of decoupling, we obtain clear single coherence dips as shown in Fig.~\ref{fig:FigSpectrumMg}
(b). The peaks caused by single spins $\text{Q}_{j}$ are indicated. A comparison
of Figs. \ref{fig:FigSpectrumMg}~(a) and (b) reveals a small peak shift of order
$\frac{1}{4}\frac{(\Omega_{j}^{\text{rfd}})^{2}}{\omega_{\text{rfd}}}\approx\frac{\Delta^{2}}{2\gamma_{j}B_{z}}\approx0.22$~kHz
which is analogous to the Bloch-Siegert shift \cite{abragam1961principles} and originates
from condition of Eq.~(\ref{eq:RWAStrongBzRF}). It should be stressed that this shift
does not affect the single spin addressing capability. Indeed Fig.~\ref{fig:FigSpectrumMg}~(b)
shows that in presence of the decoupling field, the signal response with and without
internuclear interaction are essentially indistinguishable.

Because of the large magnetic field in this case, $\omega_{j}\approx |\gamma_{j} B_{z}-\frac{m_{s}}{2}A_{j}^{\parallel}|$
and we can write $\boldsymbol{A}_{j} \approx A_{j}^{\parallel}\hat{z} + A_{j}^{\perp}\hat{\theta}_j$,
where the unit vector $\hat{\theta}_{j}$ is perpendicular to $\hat{z}$. We can obtain $A_{j}^{\parallel}$
with high accuracy using only the spectrum shown Fig.~\ref{fig:FigSpectrumMg}~(b). Under a rf decoupling
field, the Hamiltonian addressing the $j$-th spin is described by Eq.~(\ref{eq:Hint_RfApm}) and it changes
the coherence as
\begin{equation}
    L_{0,m_{s}}(t)=L_{0,m_{s}}^{j}(t)\equiv\cos\left(\frac{1}{8} f_{k_{\text{DD}}} |\boldsymbol{a}_{j}^{\pm}| t \right), \label{eq:Lcoher_apm}
\end{equation}
where the coupling strength $|\boldsymbol{a}_{j}^{(\pm)}|$ is proportional to $|\boldsymbol{A}_{j}^{x}|\approx  A_{j}^{\perp}$
[see Eq.~(\ref{eq:Abs:apm})]. Eq.~(\ref{eq:Lcoher_apm}) enables measurements of $A_{j}^{\perp}$.

\begin{figure}
\includegraphics[width=0.95\columnwidth]{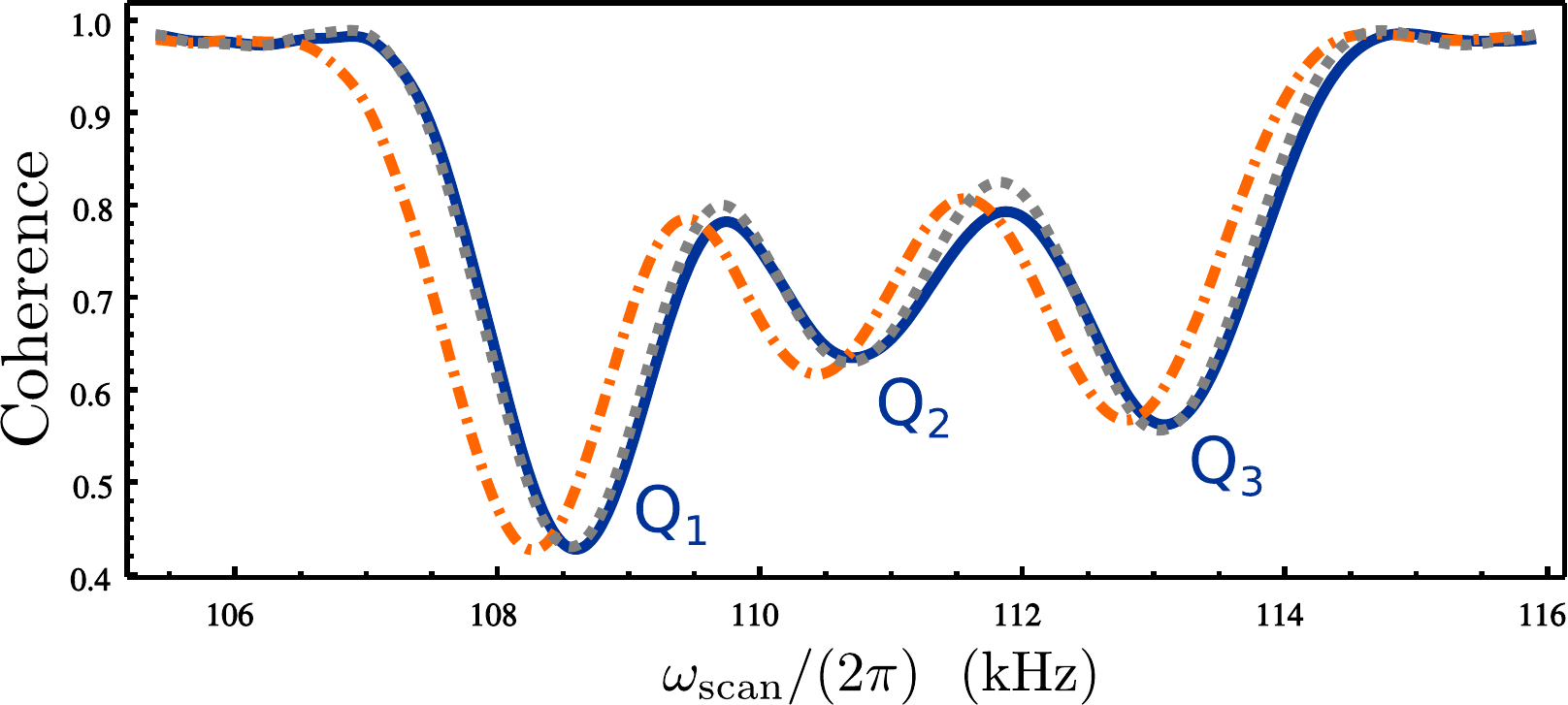}\caption{\label{fig:FigSign}(color online).
Coherence signal of NV electron spin (the gray dotted line)
modulated by the spin cluster denoted in Fig.~\ref{fig:FigSpectrumMg} and
the calculated ones using the measured parameters. The
coherence calculated by the right choice of the measured parameters
(the blue solid line) reproduces the actual one (the gray dotted line).  Changing the measured
values $\hat{\theta}_{j}$ to $-\hat{\theta}_{j}$
gives a calculated coherence (the orange dashed dotted line) that is
different from the actual one. By comparing the actual coherence and the calculated ones, the directions
of the hyperfine fields and hence the nuclear positions of the spin cluster are completely determined.
}
\end{figure}

We then apply a rf control field and obtain two possible horizontal directions
$\pm\hat{\theta}_{j}$ by solving Eq.~(\ref{eq:parallel}). From Fig.~\ref{fig:FigSpectrumMg}
we know that the carbon atoms in the spin cluster are closely located. Without
determining the sign of $\pm\hat{\theta}_j$, we find two possible positions of
the spin cluster (the physical one with $\hat{\theta}_{j}$ and its rotated image
with $\hat{\theta}_{j}\rightarrow -\hat{\theta}_{j}$). The measured results already
provide complete information on the cluster structure (relative positions of nuclear
spins). To distinguish the actual position relative to the NV center, we compare the
calculated coherence signals with the actual one in Fig.~\ref{fig:FigSign}, from
which we rule out the non-physical one. The measured positions of the nuclear spins
are listed in Table~\ref{table:clusterPositions}, from which we can see that the
errors are much smaller than the smallest distance between the carbon nuclei $a_{nn}$
in the diamond lattice.

\subsection{3D imaging for bio-molecules}\label{malate}
In this section we demonstrate that our method applies to the determination of
the molecular structures. To this end we consider as an example
of malic acid (molecular formula $\text{C}_{4}\text{H}_{6}\text{O}_{5}$) which
plays an important role in biochemistry and is not too large to prevent exact numerical simulations. Malic acid has two stereoisomeric forms,
L- and D-enantiomers. The L-malic acid is the naturally occurring form, while the
structure of D-malic is the non-superimposable mirror reflection image of L-malic.
Discriminating between the L and D-form typically requires macroscopic samples to
allow for example the polarisation rotation of light. Here we aim to show that
employing the methods that have been developed here it would become possible
to achieve this task on the level of individual molecules. We perform the simulation
for a molecule of L-malic acid placed on the surface of a diamond, and an NV center
that is implanted 2 nm below the diamond surface. The molecule consists of four
$^{12}\text{C}$ (natural abundance 98.89\%), six $^{1}\text{H}$ (natural abundance 99.985\%),
and five $^{16}\text{O}$ (natural abundance 99.76\%). The six $^{1}\text{H}$ spins in
the molecule couple to the NV electron spin, and their locations are listed in Table~\ref{table:MalicStructue}.

\begin{table}[]
\centering
\caption{
Positions of the $^{1}\text{H}$ spins in a molecule of L-malic acid
used in the simulations, as well as the measured values.
The direction {[}1,1,1{]} is chosen as the NV symmetry axis
and the NV center is located at {[}0,0,0{]}.}
\label{table:MalicStructue}
\begin{tabular}{l|ccc|ccc}
  			 & \multicolumn{3}{c|}{Positions of  $^{1}\text{H}$ spins (nm)} & \multicolumn{3}{c}{Measured values (nm)} \\ \hline
H1   & {[}1.5422, & 1.2168, & 0.4548{]} & {[}1.53, & 1.24, & 0.45{]} \\
H2   & {[}1.7412, & 1.3801, & 0.6257{]} & {[}1.78, & 1.33, & 0.61{]}  \\
H3   & {[}1.5652, & 1.4078, & 0.6208{]} & {[}1.58, & 1.40, & 0.62{]}   \\
H4   & {[}1.8230, & 1.2016, & 0.4986{]}  & {[}1.78, & 1.29, & 0.50{]} \\
H5   & {[}1.4962, & 1.1024, & 0.7883{]} & {[}1.48, & 1.14, & 0.78{]}  \\
H6   & {[}1.7007, & 1.6559, & 0.4148{]} & {[}1.74, & 1.61, & 0.41{]}
\end{tabular}
\end{table}

\begin{figure}
\includegraphics[width=0.95\columnwidth]{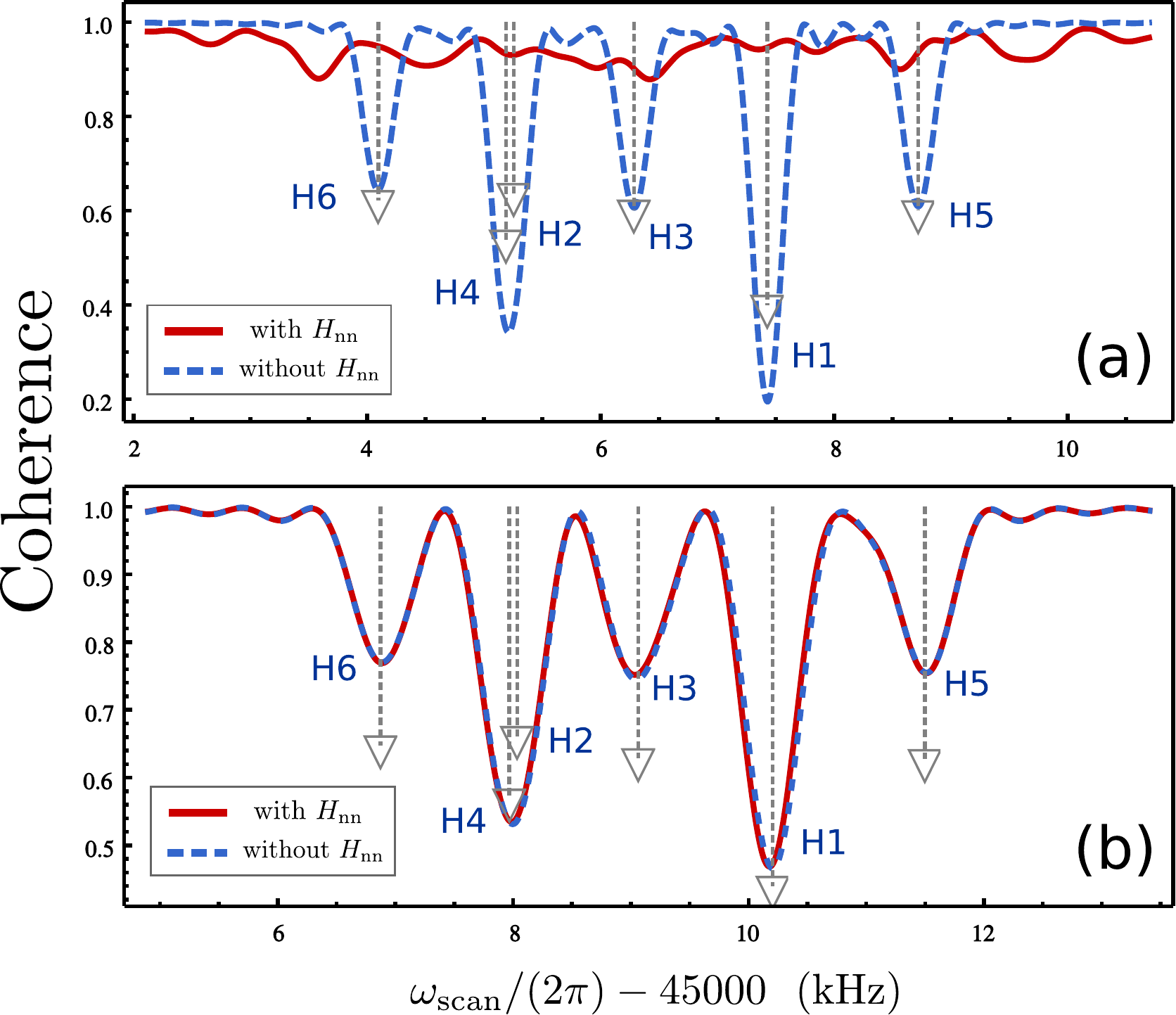}\caption{\label{fig:FigSpectrumMol}(color online).
Coherence signals between the NV $|-1\rangle$ and $|0\rangle$ states modulated by a L-malic acid
molecule (see Table~\ref{table:MalicStructue} for spin positions).
A magnetic field $B_{z}\approx 1.057$ T
is applied along the NV symmetry axis and we choose $f_{k_{\text{DD}}}=0.028$.
The vertical down arrows indicate the locations of single spin
resonant frequencies, with the lengths denote the relative coupling
strengths. The red solid line in (a) is the case without rf decoupling
field, while the red line in (b) is the case that a rf decoupling
field with $\Delta=2\pi\times 500$ kHz is applied to suppress nuclear
dipolar coupling. The blue dashed lines in (a) and
(b) are the corresponding simulations without considering nuclear
dipolar interaction. The application of rf decoupling reduces the electron-nuclear interactions by a factor $1/\sqrt{3}$ and hence reduces the separation between resonances.}
\end{figure}

We employ the methods described in the previous subsection~\ref{subsec:C13Cluster}
to determine the nuclear positions. In the simulation, we apply a strong magnetic field
$B_{z}\approx 1.057$ T. The AXY sequences utilize a high harmonic $k_{\text{DD}}=225$ to
reduce the total number of composite pulses to $1200$, which corresponds to a total evolution
time of around 3 ms. Compared with the case without rf decoupling in Fig.~\ref{fig:FigSpectrumMol} (a),
our method reveals the resonant structures [see Fig.~\ref{fig:FigSpectrumMol} (b)] and
enable individual measurements of the nuclear spins in the molecule. It can be seen in
Fig.~\ref{fig:FigSpectrumMol} that with the nuclear spins decoupled the signal of each
spin resonance is more pronounced. The spins labeled by $\text{H2}$ and $\text{H4}$ are
indistinguishable in the spectrum shown in Fig.~\ref{fig:FigSpectrumMol} (b). But since
the spins are decoupled, we can identify from the coherence patterns at different control
parameters that there are two addressed spins with approximately the same $A_{j}^{\parallel}$.
Using a model of two nuclear spins, we obtain $A_{j}^{\perp}$ by fitting
$\prod_{j=\text{H2},\text{H4}} L_{0,m_{s}}^{j}(t)$ [see Eq.~(\ref{eq:Lcoher_apm})] with
the coherence data and infer the directions by adding a rf control field.
The measured positions of all the spins in the bio-molecule are listed in Table~\ref{table:MalicStructue}.
With the measured positions, we can identify that the malic acid is the L-enantiomer.
With the resolved structure of the spins, the accuracy of the measured positions
can be further improved if we compare the calculated and actual coherence data using
a larger $f_{k_{\text{DD}}}$ and (or) keeping nuclear dipolar coupling.

We can estimate the average time we need to resolve each individual peak in Fig.~\ref{fig:FigSpectrumMol} as follows. Assuming a signal of $\sim 0.1$ (the signals in Fig.~\ref{fig:FigSpectrumMol} (b) are larger), we need 100 measurements to reach the signal-to-noise threshold. Using a collection efficiency of $0.2$ (see~\cite{ajoy2015atomic}), the time per shot equalling to the evolution time $\sim 3$ ms (the time for NV initialization and measurement is a few microseconds and therefore negligible in the estimation), and a sampling of $10$ points per peak, we find that a time of $\sim 15$ seconds would be sufficient to resolve each peak. Similarly, the time to determine the relative direction of one spin is $\sim 15$ seconds. In this manner, the extraction of each relative spin position takes  $\sim 0.5$ minutes.

Finally, we want to comment that the noise induced by the surface proximity has been identified as a possible limiting factor
 for any method trying to obtain information about external spins. More specifically, in Ref.~\cite{romach2015spectroscopy} that noise is determined
 as originating from two different sources, that is slow noise generated by a surface electronic spin bath and a faster noise attributed 
 to surface-modified phononic coupling.  In this respect we want to comment that  while the slow frequency noise can be significantly mitigated by  the application of a high number of microwave pulses, note that in our case the AXY sequence employs $1200$ composite pulses, the faster noise can be reduced when the system is driven towards lower temperatures. 
Additionally we want to comment that further experimental developments as the one presented in Ref.~\cite{lovchinsky2016nuclear} have significantly enhanced the coherence times of shallow NV centers. 
Furthermore, we want to remark that our method is equally useful for shorter DD sequences, i.e., when the sensing time of each run is reduced, and$/$or when applied  on NV centers located at a larger distance from the surface.  
In these cases, the signal peaks have overlaps. However, the reconstruction of spin clusters can be efficiently assisted by numerical methods because of the presence of the decoupling field that allows to deal the measured signals as coming from non-interacting spins~\cite{zhao2012sensing}.

\section{Summary and discussions \label{sec:Summary}}
We have developed methods to individually address and control interacting
nuclear spins in ensembles through a single NV electron spin. Our protocol
allows  the suppression of unwanted nucleus-nucleus coupling and interactions
between electron spin and the background noise. Thereby the hyperfine fields
and hence the 3D positions of nuclear spins are measured. The interactions between
the center electron spins and the addressed nuclear spins can be tuned continuously
and are changeable to different types of interaction Hamiltonians. Therefore,
high-fidelity two-qubit quantum gates between NV electron spin and nuclear
spins or selective spin polarization can be implemented by our techniques.
With the measurement of the nuclear positions of interacting spins, the structure
of the spin clusters can be identified and the nuclear dipole-dipole interaction
can be obtained. We applied the method to the measured the positions of
all the proton spins in a molecule of L-malic acid to show the principles at work.
The measured positions are sufficiently accurate to identify the stereoisomeric
form of the molecule of malic acid. Our work paves the way to resolve and measure
the structures of large-size nuclear spin clusters, which is ideal for solid-state quantum
simulators with a long coherence time~\cite{cai2013large}. The theory and methods
developed are not restricted to the study of nuclear spins and  can be applied
to other spin systems, e.g., systems of an NV center coupled to electron spin
labels in bio-molecules.

\section*{Acknowledgements}
This work was supported by the Alexander von Humboldt Foundation, the ERC
Synergy grant BioQ, the EU projects DIADEMS, SIQS and EQUAM as well as the
DFG via the SFB TRR/21 and the SPP 1601. Simulations were performed on the
computational resource bwUniCluster funded by the Ministry of Science, Research
and the Arts Baden-W\"urttemberg and the Universities of the State of
Baden-W\"urttemberg, Germany, within the framework program bwHPC. 
We thank Fedor Jelezko and Thomas Unden for discussions.

Z.-Y. W. and J. C. contributed equally to this work.

\appendix

\section{Some details of the theory\label{sec:Appendix}}

\subsection{Derivation for $\tilde{\boldsymbol{A}}_{j}(t)$}

We calculate the field $\tilde{\boldsymbol{A}}_{j}(t)$ in Eq.~(\ref{eq:Hint}) of the main text. $\tilde{\boldsymbol{A}}_{j}(t)\cdot\boldsymbol{I}_{j}=U_{\text{n}}^{\dagger}(t)\boldsymbol{A}_{j}\cdot\boldsymbol{I}_{j}U_{\text{n}}(t)$,
where the unitary operator $U_{\text{n}}(t)=\mathcal{T}e^{-i\int_{0}^{t}H_{\text{n}}(\tau)d\tau}$
is governed by the Hamiltonian $H_{\text{n}}(t)= H_{\text{nZ}}^{\text{eff}}+H_{\text{rfd}}(t)$.
We write $U_{\text{n}}(t)$ in the rotating frame with respect to
$ H_{\text{nZ}}^{\text{eff}}$,
\begin{equation}
    U_{\text{n}}(t)=U_{\text{n}Z}(t)\tilde{U}_{\text{rfd}}(t),\label{eq:Unj1}
\end{equation}
where $U_{\text{n}Z}(t)=e^{-i H_{\text{nZ}}^{\text{eff}}t}=e^{i\sum_{j}\boldsymbol{\omega}_{j}\cdot\boldsymbol{I}_{j}t}$
and $\tilde{U}_{\text{rfd}}(t)=\mathcal{T}e^{-i\int_{0}^{t}\tilde{H}_{\text{rfd}}(\tau)d\tau}$
with $\tilde{H}_{\text{rfd}}(t)=U_{nZ}^{\dagger}(t)H_{\text{rfd}}(t)U_{nZ}(t)$.
We obtain
\begin{eqnarray}
    \tilde{H}_{\text{rfd}}(t) & = & \sum_{j}\gamma_{j}V_{\text{rfd}}\cos(\omega_{\text{rfd}}t-\phi_{\text{rfd}})\tilde{\boldsymbol{n}}_{j}(t)\cdot\boldsymbol{I}_{j},
\end{eqnarray}
where
\begin{equation}
    \tilde{\boldsymbol{n}}_{j}(t)\cdot\boldsymbol{I}_{j}=e^{-i\boldsymbol{\omega}_{j}\cdot\boldsymbol{I}_{j}t}\hat{n}_{\text{rf}}\cdot\boldsymbol{I}_{j}e^{i\boldsymbol{\omega}_{j}\cdot\boldsymbol{I}_{j}t}.
\end{equation}
With the aid of the identity
\begin{equation}
    e^{i\boldsymbol{I}_{j}\cdot\hat{l}\phi}\boldsymbol{I}_{j}\cdot\boldsymbol{b}e^{-i\boldsymbol{I}_{j}\cdot\hat{l}\phi}=\boldsymbol{I}_{j}\cdot[(\boldsymbol{b}-\boldsymbol{b}\cdot\hat{l}\hat{l})\cos\phi-\hat{l}\times\boldsymbol{b}\sin\phi+\boldsymbol{b}\cdot\hat{l}\hat{l}],\label{eq:formula}
\end{equation}
which rotates the field $\boldsymbol{b}$ around the axis $\hat{l}$ by an angle $\phi$, we get
\begin{eqnarray}
\tilde{\boldsymbol{n}}_{j}(t) & = & \boldsymbol{n}_{j}^{x}\cos\omega_{j}t+\boldsymbol{n}_{j}^{y}\sin\omega_{j}t+\boldsymbol{n}_{j}^{z},\\
\boldsymbol{n}_{j}^{x} & = & \hat{n}_{\text{rf}}-\hat{\omega}_{j}(\hat{n}_{\text{rf}}\cdot\hat{\omega}_{j}),\\
\boldsymbol{n}_{j}^{y} & = & \hat{\omega}_{j}\times\hat{n}_{\text{rf}},\\
\boldsymbol{n}_{j}^{z} & = & \hat{n}_{\text{rf}}\cdot\hat{\omega}_{j}\hat{\omega}_{j}.
\end{eqnarray}
For the rf driving at the frequency   $\omega_{\text{rf}d}\sim\omega_{j}\gg\gamma_{j}V_{\text{rfc}}$,
we apply the rotating wave approximation and get
\begin{equation}
\tilde{H}_{\text{rfd}}(t)=\sum_{j}\frac{\gamma_{j}V_{\text{rfd}}}{2}\left[\boldsymbol{n}_{j}^{x}\cos\varphi_{j}^{\text{rfd}}(t)+\boldsymbol{n}_{j}^{y}\sin\varphi_{j}^{\text{rfd}}(t)\right]\cdot\boldsymbol{I}_{j},
\end{equation}
with $\varphi_{j}^{\text{rfd}}(t)=\phi_{\text{rfd}}-(\omega_{\text{rfd}}-\omega_{j})t$.
The time dependence in $\tilde{H}_{\text{rfd}}(t)$ can be removed
in a frame $e^{i\sum_{j}(\omega_{\text{rfd}}-\omega_{j})\hat{\omega}_{j}\cdot\boldsymbol{I}_{j}t}$.
That is,
\begin{equation}
\tilde{U}_{\text{rfd}}(t)=e^{i\sum_{j}(\omega_{\text{rfd}}-\omega_{j})\hat{\omega}_{j}\cdot\boldsymbol{I}_{j}t}e^{-iH_{\nu}t},\label{eq:UrfTilde}
\end{equation}
where the Hamiltonian
\begin{eqnarray}
H_{\nu} & = & \sum_{j}\boldsymbol{\nu}_{j}\cdot\boldsymbol{I}_{j},\\
\boldsymbol{\nu}_{j} & = & \frac{1}{2}\gamma_{j}V_{\text{rfd}}\boldsymbol{n}_{j}({\phi_{\text{rfd}}})+(\omega_{\text{rfd}}-\omega_{j})\hat{\omega}_{j},
\end{eqnarray}
with $\boldsymbol{\nu}_{j}$ given by Eq.~(\ref{eq:nuj}).
The magnitude of $\boldsymbol{\nu}_{j}\equiv\nu_{j}\hat{\nu}_{j}$ is
$\nu_{j}=\sqrt{(\Omega_{j}^{\text{rfd}})^{2}+(\omega_{\text{rfd}}-\omega_{j})^{2}}$,
where the Rabi frequency on the nuclear spin $\Omega_{j}^{\text{rfd}}=\frac{1}{2}\gamma_{j}V_{\text{rfd}}|\boldsymbol{n}_{j}^{\phi_{\text{rfd}}}|$
and $|\boldsymbol{n}_{j}^{\phi_{\text{rfd}}}|=\sqrt{1-|\hat{n}_{\text{rf}}\cdot\hat{\omega}_{j}|^{2}}$.
Under a strong magnetic field $\hat{\omega}_j \approx \hat{z}$  and
$|\boldsymbol{n}_{j}^{\phi_{\text{rfd}}}|\approx\sqrt{1-|\hat{n}_{\text{rf}}\cdot\hat{z}|^{2}}$.
Summarizing Eqs.~(\ref{eq:Unj1}) and (\ref{eq:UrfTilde}), we have,
under the condition $\omega_{\text{rfd}}\sim\omega_{j}\gg\gamma_{j}V_{\text{rfd}}$,
\begin{equation}
U_{\text{n}}(t)=U_{\omega}(t)U_{\nu}(t),\label{eq:Un}
\end{equation}
where $U_{\omega,\nu}(t)=e^{-iH_{\omega,\nu}t}$ with
\begin{equation}
H_{\omega}=-\omega_{\text{rfd}}\sum_{j}\hat{\omega}_{j}\cdot\boldsymbol{I}_{j}.
\end{equation}
Using Eqs.~(\ref{eq:Un}) and (\ref{eq:formula}), we obtain Eq.~(\ref{eq:AjTilde}) for the field $\tilde{\boldsymbol{A}}_{j}(t)$.

\subsection{Decoupling of nucleus-nucleus interaction}

The internuclear interaction is $H_{\text{nn}}=\sum_{j>k}\frac{\mu_{0}}{4\pi}\frac{\gamma_{j}\gamma_{k}}{r_{j,k}^{3}}D_{(j,k)}$,
where $r_{j,k}$ is the distance between the nuclear spins located
at $\boldsymbol{r}_{j}$ and $\boldsymbol{r}_{k}$. The operator

\begin{equation}
D_{(j,k)}=\boldsymbol{I}_{j}\cdot\boldsymbol{I}_{k}-3(\boldsymbol{I}_{j}\cdot\hat{r}_{j,k})(\hat{r}_{j,k}\cdot\boldsymbol{I}_{k}),
\end{equation}
where $\hat{r}_{j,k}$ is the unit vector of $\boldsymbol{r}_{j}-\boldsymbol{r}_{k}$.
Here we show that, when dealing  with the same nuclear specie ($\gamma_{j}\approx\gamma_{k}$), 
$H_{\text{nn}}$ can be suppressed by the nuclear Hamiltonian $H_{\text{n}}(t)$ even though the nuclear
spins feel different Larmor frequency $\omega_{j}$. Because $H_{\text{nn}}$
is expressed  by a linear combination of the operators $D_{(j,k)}$, it is sufficient
to see the decoupling for two nuclear spins.

In the interaction picture of $H_{\text{n}}(t)$, $\tilde{D}_{(j,k)}(t)=U_{\text{n}}^{\dagger}(t)D_{(j,k)}U_{\text{n}}(t)=U_{\nu}^{\dagger}(t)\tilde{D}_{(j,k)}^{(1)}(t)U_{\nu}(t)$,
where $\tilde{D}_{(j,k)}^{(1)}(t)=U_{\omega}^{\dagger}(t)D_{(j,k)}U_{\omega}(t)$
is the operator written in the interaction picture of the Hamiltonian
$H_{\omega}$. For the Hamiltonian $H_{\omega}$, each of the spins has
an energy $\omega_{\text{rfd}}$ much larger than the dipolar coupling.
Therefore the terms in $D_{(j,k)}$ that do not conserve the energy
carry fast oscillating factors in $\tilde{D}_{(j,k)}^{(1)}(t)$ and
are suppressed by the large $\omega_{\text{rfd}}$. Neglecting those
terms that do not conserve the energy by the secular approximation
(or equivalently by the rotating wave approximation), we have
\begin{eqnarray}
\tilde{D}_{(j,k)}^{(1)}(t) & \cong & \frac{1}{2}[1-3(\hat{r}_{j,k}\cdot\hat{z})^{2}][3I_{j}^{z}I_{k}^{z}-\boldsymbol{I}_{j}\cdot\boldsymbol{I}_{k}],\label{eq:D1jk}
\end{eqnarray}
where $\cong$ means that the non-secular terms have been dropped
out. That is, we have
\begin{equation}
H_{\text{nn}}\approx\sum_{j>k}\frac{\mu_{0}}{8\pi}\frac{\gamma_{j}\gamma_{k}}{r_{j,k}^{3}}\left[1-3(\hat{r}_{j,k}\cdot\hat{z})^{2}\right]\left(3I_{j}^{z}I_{k}^{z}-\boldsymbol{I}_{j}\cdot\boldsymbol{I}_{k}\right).
\end{equation}

We consider the regime of strong magnetic field that $|\gamma_{j}B_{z}|\gg|\boldsymbol{A}_{j}|$.
This allows the simplification that the unit vector $\hat{\omega}_{j}\approx\hat{z}$
and the Rabi frequency $\Omega_{j}^{\text{rfd}}\approx\sqrt{2}\Delta$
are approximately the same. We define the coordinates $\hat{z}=\hat{\omega}_{j}$,
$\hat{x}=\boldsymbol{n}_{j}({\phi_{\text{rfd}}})/|\boldsymbol{n}_{j}({\phi_{\text{rfc}}})|$,
and $\hat{y}=\hat{z}\times\hat{x}$. In this coordinate, $\boldsymbol{\nu}_{j}=\sqrt{2}\Delta\hat{x}+(\Delta+\delta_{j})\hat{z}$
with $\delta_{j}$ given by Eq.~(\ref{eq:deltaj}). If $\delta_{j}=0$  the coupling operator $\tilde{D}_{(j,k)}^{(1)}(t)$ in Eq.
(\ref{eq:D1jk}) can be further suppressed by the magic spinning.
Here we show how to suppress the dipolar coupling in the realistic
case that $\delta_{j}\neq0$. Let a new coordinate $\hat{z}_{j}=\hat{\nu}_{j}$,
$\hat{y}_{j}=\hat{y}$, and $\hat{x}_{j}=\hat{y}_{j}\times\hat{z}_{j}$, and define $I_{j}^{\alpha_{j}}=\boldsymbol{I}_{j}\cdot\hat{\alpha}_{j}$.
We chose the detuning on the target coupled nuclei and driving amplitude
that $\delta_{j}/\Delta\ll1$ and define $\eta=(\delta_{j}+\delta_{k})/\Delta$.
We obtain
\begin{eqnarray*}
2I_{j}^{z}I_{k}^{z} & \cong & \frac{2}{3}I_{j}^{z_{j}}I_{k}^{z_{k}}(1+\frac{2}{3}\eta)+\frac{4}{3}I_{j}^{x_{j}}I_{k}^{x_{k}}(1-\frac{1}{3}\eta)+O(\eta^{2}), \\
I_{j}^{x}I_{k}^{x} & \cong & \frac{2}{3}I_{j}^{z_{j}}I_{k}^{z_{k}}(1-\frac{1}{3}\eta)+\frac{1}{3}I_{j}^{x_{j}}I_{k}^{x_{k}}(1+\frac{2}{3}\eta)+O(\eta^{2}), \\
I_{j}^{y}I_{k}^{y} & \cong & I_{j}^{y_{j}}I_{k}^{y_{k}},
\end{eqnarray*}
which implies from Eq.~(\ref{eq:D1jk}) that after rf decoupling $\tilde{D}_{(j,k)}^{(1)}(t)\cong O(\eta)$
has residual nuclear coupling because of $\delta_{j}\neq0$. The residual coupling is suppressed by the
reduction factor $\eta$. For the addressing of the $k$-th spin, we choose $\delta_k=0$ and nuclear dipolar
coupling to the addressed spin is negligible as long as the reduction factor $\eta=\delta_{j}/\Delta\ll1$.


\begin{thebibliography}{42}%
\makeatletter
\providecommand \@ifxundefined [1]{%
 \@ifx{#1\undefined}
}%
\providecommand \@ifnum [1]{%
 \ifnum #1\expandafter \@firstoftwo
 \else \expandafter \@secondoftwo
 \fi
}%
\providecommand \@ifx [1]{%
 \ifx #1\expandafter \@firstoftwo
 \else \expandafter \@secondoftwo
 \fi
}%
\providecommand \natexlab [1]{#1}%
\providecommand \enquote  [1]{``#1''}%
\providecommand \bibnamefont  [1]{#1}%
\providecommand \bibfnamefont [1]{#1}%
\providecommand \citenamefont [1]{#1}%
\providecommand \href@noop [0]{\@secondoftwo}%
\providecommand \href [0]{\begingroup \@sanitize@url \@href}%
\providecommand \@href[1]{\@@startlink{#1}\@@href}%
\providecommand \@@href[1]{\endgroup#1\@@endlink}%
\providecommand \@sanitize@url [0]{\catcode `\\12\catcode `\$12\catcode
  `\&12\catcode `\#12\catcode `\^12\catcode `\_12\catcode `\%12\relax}%
\providecommand \@@startlink[1]{}%
\providecommand \@@endlink[0]{}%
\providecommand \url  [0]{\begingroup\@sanitize@url \@url }%
\providecommand \@url [1]{\endgroup\@href {#1}{\urlprefix }}%
\providecommand \urlprefix  [0]{URL }%
\providecommand \Eprint [0]{\href }%
\providecommand \doibase [0]{http://dx.doi.org/}%
\providecommand \selectlanguage [0]{\@gobble}%
\providecommand \bibinfo  [0]{\@secondoftwo}%
\providecommand \bibfield  [0]{\@secondoftwo}%
\providecommand \translation [1]{[#1]}%
\providecommand \BibitemOpen [0]{}%
\providecommand \bibitemStop [0]{}%
\providecommand \bibitemNoStop [0]{.\EOS\space}%
\providecommand \EOS [0]{\spacefactor3000\relax}%
\providecommand \BibitemShut  [1]{\csname bibitem#1\endcsname}%
\let\auto@bib@innerbib\@empty
\bibitem [{\citenamefont {Doherty}\ \emph {et~al.}(2013)\citenamefont
  {Doherty}, \citenamefont {Manson}, \citenamefont {Delaney}, \citenamefont
  {Jelezko}, \citenamefont {Wrachtrup},\ and\ \citenamefont
  {Hollenberg}}]{doherty2013nitrogen}%
  \BibitemOpen
  \bibfield  {author} {\bibinfo {author} {\bibfnamefont {M.~W.}\ \bibnamefont
  {Doherty}}, \bibinfo {author} {\bibfnamefont {N.~B.}\ \bibnamefont {Manson}},
  \bibinfo {author} {\bibfnamefont {P.}~\bibnamefont {Delaney}}, \bibinfo
  {author} {\bibfnamefont {F.}~\bibnamefont {Jelezko}}, \bibinfo {author}
  {\bibfnamefont {J.}~\bibnamefont {Wrachtrup}}, \ and\ \bibinfo {author}
  {\bibfnamefont {L.~C.~L.}\ \bibnamefont {Hollenberg}},\ }\bibfield  {title}
  {\enquote {\bibinfo {title} {The nitrogen-vacancy colour centre in
  diamond},}\ }\href@noop {} {\bibfield  {journal} {\bibinfo  {journal} {Phys.
  Rep.}\ }\textbf {\bibinfo {volume} {528}},\ \bibinfo {pages} {1} (\bibinfo
  {year} {2013})}\BibitemShut {NoStop}%
\bibitem [{\citenamefont {Dobrovitski}\ \emph {et~al.}(2013)\citenamefont
  {Dobrovitski}, \citenamefont {Fuchs}, \citenamefont {Falk}, \citenamefont
  {Santori},\ and\ \citenamefont {Awschalom}}]{dobrovitski2013quantum}%
  \BibitemOpen
  \bibfield  {author} {\bibinfo {author} {\bibfnamefont {V.~V.}\ \bibnamefont
  {Dobrovitski}}, \bibinfo {author} {\bibfnamefont {G.~D.}\ \bibnamefont
  {Fuchs}}, \bibinfo {author} {\bibfnamefont {A.~L.}\ \bibnamefont {Falk}},
  \bibinfo {author} {\bibfnamefont {C.}~\bibnamefont {Santori}}, \ and\
  \bibinfo {author} {\bibfnamefont {D.~D.}\ \bibnamefont {Awschalom}},\
  }\bibfield  {title} {\enquote {\bibinfo {title} {Quantum control over single
  spins in diamond},}\ }\href@noop {} {\bibfield  {journal} {\bibinfo
  {journal} {Annu. Rev. Condens. Matter Phys.}\ }\textbf {\bibinfo {volume}
  {4}},\ \bibinfo {pages} {23} (\bibinfo {year} {2013})}\BibitemShut {NoStop}%
\bibitem [{\citenamefont {Wu}\ \emph {et~al.}(2015)\citenamefont {Wu},
  \citenamefont {Jelezko}, \citenamefont {Plenio},\ and\ \citenamefont
  {Weil}}]{WuJPW15}%
  \BibitemOpen
  \bibfield  {author} {\bibinfo {author} {\bibfnamefont {Y.}~\bibnamefont
  {Wu}}, \bibinfo {author} {\bibfnamefont {F.}~\bibnamefont {Jelezko}},
  \bibinfo {author} {\bibfnamefont {M.~B.}\ \bibnamefont {Plenio}}, \ and\
  \bibinfo {author} {\bibfnamefont {T.}~\bibnamefont {Weil}},\ }\bibfield
  {title} {\enquote {\bibinfo {title} {Diamond quantum devices in biology},}\
  }\href@noop {} {\bibfield  {journal} {\bibinfo  {journal} {Submitted to
  Angew. Chemie}\ } (\bibinfo {year} {2015})}\BibitemShut {NoStop}%
\bibitem [{\citenamefont {Zhao}\ \emph {et~al.}(2011)\citenamefont {Zhao},
  \citenamefont {Hu}, \citenamefont {Ho}, \citenamefont {Wan},\ and\
  \citenamefont {Liu}}]{zhao2011atomic}%
  \BibitemOpen
  \bibfield  {author} {\bibinfo {author} {\bibfnamefont {N.}~\bibnamefont
  {Zhao}}, \bibinfo {author} {\bibfnamefont {J.-L.}\ \bibnamefont {Hu}},
  \bibinfo {author} {\bibfnamefont {S.-W.}\ \bibnamefont {Ho}}, \bibinfo
  {author} {\bibfnamefont {J.~T.~K.}\ \bibnamefont {Wan}}, \ and\ \bibinfo
  {author} {\bibfnamefont {R.-B.}\ \bibnamefont {Liu}},\ }\bibfield  {title}
  {\enquote {\bibinfo {title} {Atomic-scale magnetometry of distant nuclear
  spin clusters via nitrogen-vacancy spin in diamond},}\ }\href@noop {}
  {\bibfield  {journal} {\bibinfo  {journal} {Nat. Nanotechnol.}\ }\textbf
  {\bibinfo {volume} {6}},\ \bibinfo {pages} {242} (\bibinfo {year}
  {2011})}\BibitemShut {NoStop}%
\bibitem [{\citenamefont {Kolkowitz}\ \emph {et~al.}(2012)\citenamefont
  {Kolkowitz}, \citenamefont {Unterreithmeier}, \citenamefont {Bennett},\ and\
  \citenamefont {Lukin}}]{kolkowitz2012sensing}%
  \BibitemOpen
  \bibfield  {author} {\bibinfo {author} {\bibfnamefont {S.}~\bibnamefont
  {Kolkowitz}}, \bibinfo {author} {\bibfnamefont {Q.~P.}\ \bibnamefont
  {Unterreithmeier}}, \bibinfo {author} {\bibfnamefont {S.~D.}\ \bibnamefont
  {Bennett}}, \ and\ \bibinfo {author} {\bibfnamefont {M.~D.}\ \bibnamefont
  {Lukin}},\ }\bibfield  {title} {\enquote {\bibinfo {title} {Sensing distant
  nuclear spins with a single electron spin},}\ }\href@noop {} {\bibfield
  {journal} {\bibinfo  {journal} {Phys. Rev. Lett.}\ }\textbf {\bibinfo
  {volume} {109}},\ \bibinfo {pages} {137601} (\bibinfo {year}
  {2012})}\BibitemShut {NoStop}%
\bibitem [{\citenamefont {Taminiau}\ \emph {et~al.}(2012)\citenamefont
  {Taminiau}, \citenamefont {Wagenaar}, \citenamefont {van~der Sar},
  \citenamefont {Jelezko}, \citenamefont {Dobrovitski},\ and\ \citenamefont
  {Hanson}}]{taminiau2012detection}%
  \BibitemOpen
  \bibfield  {author} {\bibinfo {author} {\bibfnamefont {T.~H.}\ \bibnamefont
  {Taminiau}}, \bibinfo {author} {\bibfnamefont {J.~J.~T.}\ \bibnamefont
  {Wagenaar}}, \bibinfo {author} {\bibfnamefont {T.}~\bibnamefont {van~der
  Sar}}, \bibinfo {author} {\bibfnamefont {F.}~\bibnamefont {Jelezko}},
  \bibinfo {author} {\bibfnamefont {V.~V.}\ \bibnamefont {Dobrovitski}}, \ and\
  \bibinfo {author} {\bibfnamefont {R.}~\bibnamefont {Hanson}},\ }\bibfield
  {title} {\enquote {\bibinfo {title} {Detection and control of individual
  nuclear spins using a weakly coupled electron spin},}\ }\href@noop {}
  {\bibfield  {journal} {\bibinfo  {journal} {Phys. Rev. Lett.}\ }\textbf
  {\bibinfo {volume} {109}},\ \bibinfo {pages} {137602} (\bibinfo {year}
  {2012})}\BibitemShut {NoStop}%
\bibitem [{\citenamefont {Zhao}\ \emph {et~al.}(2012)\citenamefont {Zhao},
  \citenamefont {Honert}, \citenamefont {Schmid}, \citenamefont {Klas},
  \citenamefont {Isoya}, \citenamefont {Markham}, \citenamefont {Twitchen},
  \citenamefont {Jelezko}, \citenamefont {Liu}, \citenamefont {Fedder},\ and\
  \citenamefont {Wrachtrup}}]{zhao2012sensing}%
  \BibitemOpen
  \bibfield  {author} {\bibinfo {author} {\bibfnamefont {N.}~\bibnamefont
  {Zhao}}, \bibinfo {author} {\bibfnamefont {J.}~\bibnamefont {Honert}},
  \bibinfo {author} {\bibfnamefont {B.}~\bibnamefont {Schmid}}, \bibinfo
  {author} {\bibfnamefont {M.}~\bibnamefont {Klas}}, \bibinfo {author}
  {\bibfnamefont {J.}~\bibnamefont {Isoya}}, \bibinfo {author} {\bibfnamefont
  {M.}~\bibnamefont {Markham}}, \bibinfo {author} {\bibfnamefont
  {D.}~\bibnamefont {Twitchen}}, \bibinfo {author} {\bibfnamefont
  {F.}~\bibnamefont {Jelezko}}, \bibinfo {author} {\bibfnamefont {R.-B.}\
  \bibnamefont {Liu}}, \bibinfo {author} {\bibfnamefont {H.}~\bibnamefont
  {Fedder}}, \ and\ \bibinfo {author} {\bibfnamefont {J.}~\bibnamefont
  {Wrachtrup}},\ }\bibfield  {title} {\enquote {\bibinfo {title} {Sensing
  single remote nuclear spins},}\ }\href@noop {} {\bibfield  {journal}
  {\bibinfo  {journal} {Nat. Nanotechnol.}\ }\textbf {\bibinfo {volume} {7}},\
  \bibinfo {pages} {657} (\bibinfo {year} {2012})}\BibitemShut {NoStop}%
\bibitem [{\citenamefont {London}\ \emph {et~al.}(2013)\citenamefont {London},
  \citenamefont {Scheuer}, \citenamefont {Cai}, \citenamefont {Schwarz},
  \citenamefont {Retzker}, \citenamefont {Plenio}, \citenamefont {Katagiri},
  \citenamefont {Teraji}, \citenamefont {Koizumi}, \citenamefont {Isoya},
  \citenamefont {Fischer}, \citenamefont {McGuinness}, \citenamefont
  {Naydenov},\ and\ \citenamefont {Jelezko}}]{london2013detecting}%
  \BibitemOpen
  \bibfield  {author} {\bibinfo {author} {\bibfnamefont {P.}~\bibnamefont
  {London}}, \bibinfo {author} {\bibfnamefont {J.}~\bibnamefont {Scheuer}},
  \bibinfo {author} {\bibfnamefont {J.-M.}\ \bibnamefont {Cai}}, \bibinfo
  {author} {\bibfnamefont {I.}~\bibnamefont {Schwarz}}, \bibinfo {author}
  {\bibfnamefont {A.}~\bibnamefont {Retzker}}, \bibinfo {author} {\bibfnamefont
  {M.~B.}\ \bibnamefont {Plenio}}, \bibinfo {author} {\bibfnamefont
  {M.}~\bibnamefont {Katagiri}}, \bibinfo {author} {\bibfnamefont
  {T.}~\bibnamefont {Teraji}}, \bibinfo {author} {\bibfnamefont
  {S.}~\bibnamefont {Koizumi}}, \bibinfo {author} {\bibfnamefont
  {J.}~\bibnamefont {Isoya}}, \bibinfo {author} {\bibfnamefont
  {R.}~\bibnamefont {Fischer}}, \bibinfo {author} {\bibfnamefont {L.~P.}\
  \bibnamefont {McGuinness}}, \bibinfo {author} {\bibfnamefont
  {B.}~\bibnamefont {Naydenov}}, \ and\ \bibinfo {author} {\bibfnamefont
  {F.}~\bibnamefont {Jelezko}},\ }\bibfield  {title} {\enquote {\bibinfo
  {title} {Detecting and polarizing nuclear spins with double resonance on a
  single electron spin},}\ }\href@noop {} {\bibfield  {journal} {\bibinfo
  {journal} {Phys. Rev. Lett.}\ }\textbf {\bibinfo {volume} {111}},\ \bibinfo
  {pages} {067601} (\bibinfo {year} {2013})}\BibitemShut {NoStop}%
\bibitem [{\citenamefont {Shi}\ \emph {et~al.}(2014)\citenamefont {Shi},
  \citenamefont {Kong}, \citenamefont {Wang}, \citenamefont {Kong},
  \citenamefont {Zhao}, \citenamefont {Liu},\ and\ \citenamefont
  {Du}}]{shi2014sensing}%
  \BibitemOpen
  \bibfield  {author} {\bibinfo {author} {\bibfnamefont {F.}~\bibnamefont
  {Shi}}, \bibinfo {author} {\bibfnamefont {X.}~\bibnamefont {Kong}}, \bibinfo
  {author} {\bibfnamefont {P.}~\bibnamefont {Wang}}, \bibinfo {author}
  {\bibfnamefont {F.}~\bibnamefont {Kong}}, \bibinfo {author} {\bibfnamefont
  {N.}~\bibnamefont {Zhao}}, \bibinfo {author} {\bibfnamefont {R.-B.}\
  \bibnamefont {Liu}}, \ and\ \bibinfo {author} {\bibfnamefont
  {J.}~\bibnamefont {Du}},\ }\bibfield  {title} {\enquote {\bibinfo {title}
  {Sensing and atomic-scale structure analysis of single nuclear-spin clusters
  in diamond},}\ }\href@noop {} {\bibfield  {journal} {\bibinfo  {journal}
  {Nat. Phys.}\ }\textbf {\bibinfo {volume} {10}},\ \bibinfo {pages} {21}
  (\bibinfo {year} {2014})}\BibitemShut {NoStop}%
\bibitem [{\citenamefont {Mkhitaryan}\ \emph {et~al.}(2015)\citenamefont
  {Mkhitaryan}, \citenamefont {Jelezko},\ and\ \citenamefont
  {Dobrovitski}}]{mkhitaryan2015highly}%
  \BibitemOpen
  \bibfield  {author} {\bibinfo {author} {\bibfnamefont {V.~V.}\ \bibnamefont
  {Mkhitaryan}}, \bibinfo {author} {\bibfnamefont {F.}~\bibnamefont {Jelezko}},
  \ and\ \bibinfo {author} {\bibfnamefont {V.~V.}\ \bibnamefont
  {Dobrovitski}},\ }\bibfield  {title} {\enquote {\bibinfo {title} {Highly
  selective detection of individual nuclear spins with rotary echo on an
  electron spin probe},}\ }\href@noop {} {\bibfield  {journal} {\bibinfo
  {journal} {Sci. Rep.}\ }\textbf {\bibinfo {volume} {5}} (\bibinfo {year}
  {2015})}\BibitemShut {NoStop}%
\bibitem [{\citenamefont {Casanova}\ \emph {et~al.}(2015)\citenamefont
  {Casanova}, \citenamefont {Wang}, \citenamefont {Haase},\ and\ \citenamefont
  {Plenio}}]{Casanova2015AXY}%
  \BibitemOpen
  \bibfield  {author} {\bibinfo {author} {\bibfnamefont {J.}~\bibnamefont
  {Casanova}}, \bibinfo {author} {\bibfnamefont {Z.-Y.}\ \bibnamefont {Wang}},
  \bibinfo {author} {\bibfnamefont {J.~F.}\ \bibnamefont {Haase}}, \ and\
  \bibinfo {author} {\bibfnamefont {M.~B.}\ \bibnamefont {Plenio}},\ }\bibfield
   {title} {\enquote {\bibinfo {title} {Robust dynamical decoupling sequences
  for individual-nuclear-spin addressing},}\ }\href@noop {} {\bibfield
  {journal} {\bibinfo  {journal} {Phys. Rev. A}\ }\textbf {\bibinfo {volume}
  {92}},\ \bibinfo {pages} {042304} (\bibinfo {year} {2015})}\BibitemShut
  {NoStop}%
\bibitem [{\citenamefont {van~der Sar}\ \emph {et~al.}(2012)\citenamefont
  {van~der Sar}, \citenamefont {Wang}, \citenamefont {Blok}, \citenamefont
  {Bernien}, \citenamefont {Taminiau}, \citenamefont {Toyli}, \citenamefont
  {Lidar}, \citenamefont {Awschalom}, \citenamefont {Hanson},\ and\
  \citenamefont {Dobrovitski}}]{van2012decoherence}%
  \BibitemOpen
  \bibfield  {author} {\bibinfo {author} {\bibfnamefont {T.}~\bibnamefont
  {van~der Sar}}, \bibinfo {author} {\bibfnamefont {Z.~H.}\ \bibnamefont
  {Wang}}, \bibinfo {author} {\bibfnamefont {M.~S.}\ \bibnamefont {Blok}},
  \bibinfo {author} {\bibfnamefont {H.}~\bibnamefont {Bernien}}, \bibinfo
  {author} {\bibfnamefont {T.~H.}\ \bibnamefont {Taminiau}}, \bibinfo {author}
  {\bibfnamefont {D.~M.}\ \bibnamefont {Toyli}}, \bibinfo {author}
  {\bibfnamefont {D.~A.}\ \bibnamefont {Lidar}}, \bibinfo {author}
  {\bibfnamefont {D.~D.}\ \bibnamefont {Awschalom}}, \bibinfo {author}
  {\bibfnamefont {R.}~\bibnamefont {Hanson}}, \ and\ \bibinfo {author}
  {\bibfnamefont {V.~V.}\ \bibnamefont {Dobrovitski}},\ }\bibfield  {title}
  {\enquote {\bibinfo {title} {Decoherence-protected quantum gates for a hybrid
  solid-state spin register},}\ }\href@noop {} {\bibfield  {journal} {\bibinfo
  {journal} {Nature}\ }\textbf {\bibinfo {volume} {484}},\ \bibinfo {pages}
  {82} (\bibinfo {year} {2012})}\BibitemShut {NoStop}%
\bibitem [{\citenamefont {Taminiau}\ \emph {et~al.}(2014)\citenamefont
  {Taminiau}, \citenamefont {Cramer}, \citenamefont {van~der Sar},
  \citenamefont {Dobrovitski},\ and\ \citenamefont
  {Hanson}}]{taminiau2014universal}%
  \BibitemOpen
  \bibfield  {author} {\bibinfo {author} {\bibfnamefont {T.~H.}\ \bibnamefont
  {Taminiau}}, \bibinfo {author} {\bibfnamefont {J.}~\bibnamefont {Cramer}},
  \bibinfo {author} {\bibfnamefont {T.}~\bibnamefont {van~der Sar}}, \bibinfo
  {author} {\bibfnamefont {V.~V.}\ \bibnamefont {Dobrovitski}}, \ and\ \bibinfo
  {author} {\bibfnamefont {R.}~\bibnamefont {Hanson}},\ }\bibfield  {title}
  {\enquote {\bibinfo {title} {Universal control and error correction in
  multi-qubit spin registers in diamond},}\ }\href@noop {} {\bibfield
  {journal} {\bibinfo  {journal} {Nat. Nanotechnol.}\ }\textbf {\bibinfo
  {volume} {9}},\ \bibinfo {pages} {171} (\bibinfo {year} {2014})}\BibitemShut
  {NoStop}%
\bibitem [{\citenamefont {Liu}\ \emph {et~al.}(2013)\citenamefont {Liu},
  \citenamefont {Po}, \citenamefont {Du}, \citenamefont {Liu},\ and\
  \citenamefont {Pan}}]{liu2013noise}%
  \BibitemOpen
  \bibfield  {author} {\bibinfo {author} {\bibfnamefont {G.-Q.}\ \bibnamefont
  {Liu}}, \bibinfo {author} {\bibfnamefont {H.~C.}\ \bibnamefont {Po}},
  \bibinfo {author} {\bibfnamefont {J.}~\bibnamefont {Du}}, \bibinfo {author}
  {\bibfnamefont {R.-B.}\ \bibnamefont {Liu}}, \ and\ \bibinfo {author}
  {\bibfnamefont {X.-Y.}\ \bibnamefont {Pan}},\ }\bibfield  {title} {\enquote
  {\bibinfo {title} {Noise-resilient quantum evolution steered by dynamical
  decoupling},}\ }\href@noop {} {\bibfield  {journal} {\bibinfo  {journal}
  {Nat. Commun.}\ }\textbf {\bibinfo {volume} {4}},\ \bibinfo {pages} {2254}
  (\bibinfo {year} {2013})}\BibitemShut {NoStop}%
\bibitem [{\citenamefont {Pfaff}\ \emph {et~al.}(2014)\citenamefont {Pfaff},
  \citenamefont {Hensen}, \citenamefont {Bernien}, \citenamefont {van Dam},
  \citenamefont {Blok}, \citenamefont {Taminiau}, \citenamefont {Tiggelman},
  \citenamefont {Schouten}, \citenamefont {Markham}, \citenamefont {Twitchen},\
  and\ \citenamefont {Hanson}}]{pfaff2014unconditional}%
  \BibitemOpen
  \bibfield  {author} {\bibinfo {author} {\bibfnamefont {W.}~\bibnamefont
  {Pfaff}}, \bibinfo {author} {\bibfnamefont {B.~J.}\ \bibnamefont {Hensen}},
  \bibinfo {author} {\bibfnamefont {H.}~\bibnamefont {Bernien}}, \bibinfo
  {author} {\bibfnamefont {S.~B.}\ \bibnamefont {van Dam}}, \bibinfo {author}
  {\bibfnamefont {M.~S.}\ \bibnamefont {Blok}}, \bibinfo {author}
  {\bibfnamefont {T.~H.}\ \bibnamefont {Taminiau}}, \bibinfo {author}
  {\bibfnamefont {M.~J.}\ \bibnamefont {Tiggelman}}, \bibinfo {author}
  {\bibfnamefont {R.~N.}\ \bibnamefont {Schouten}}, \bibinfo {author}
  {\bibfnamefont {M.}~\bibnamefont {Markham}}, \bibinfo {author} {\bibfnamefont
  {D.~J.}\ \bibnamefont {Twitchen}}, \ and\ \bibinfo {author} {\bibfnamefont
  {R.}~\bibnamefont {Hanson}},\ }\bibfield  {title} {\enquote {\bibinfo {title}
  {Unconditional quantum teleportation between distant solid-state quantum
  bits},}\ }\href@noop {} {\bibfield  {journal} {\bibinfo  {journal} {Science}\
  }\textbf {\bibinfo {volume} {345}},\ \bibinfo {pages} {532} (\bibinfo {year}
  {2014})}\BibitemShut {NoStop}%
\bibitem [{\citenamefont {M{\"u}ller}\ \emph {et~al.}(2014)\citenamefont
  {M{\"u}ller}, \citenamefont {Kong}, \citenamefont {Cai}, \citenamefont
  {Melentijevi{\'c}}, \citenamefont {Stacey}, \citenamefont {Markham},
  \citenamefont {Twitchen}, \citenamefont {Isoya}, \citenamefont {Pezzagna},
  \citenamefont {Meijer}, \citenamefont {Du}, \citenamefont {Plenio},
  \citenamefont {Naydenov}, \citenamefont {{McGuinness}},\ and\ \citenamefont
  {Jelezko}}]{muller2014nuclear}%
  \BibitemOpen
  \bibfield  {author} {\bibinfo {author} {\bibfnamefont {C.}~\bibnamefont
  {M{\"u}ller}}, \bibinfo {author} {\bibfnamefont {X.}~\bibnamefont {Kong}},
  \bibinfo {author} {\bibfnamefont {J.}~\bibnamefont {Cai}}, \bibinfo {author}
  {\bibfnamefont {K.}~\bibnamefont {Melentijevi{\'c}}}, \bibinfo {author}
  {\bibfnamefont {A.}~\bibnamefont {Stacey}}, \bibinfo {author} {\bibfnamefont
  {M.}~\bibnamefont {Markham}}, \bibinfo {author} {\bibfnamefont
  {D.}~\bibnamefont {Twitchen}}, \bibinfo {author} {\bibfnamefont
  {J.}~\bibnamefont {Isoya}}, \bibinfo {author} {\bibfnamefont
  {S.}~\bibnamefont {Pezzagna}}, \bibinfo {author} {\bibfnamefont
  {J.}~\bibnamefont {Meijer}}, \bibinfo {author} {\bibfnamefont {J.~F.}\
  \bibnamefont {Du}}, \bibinfo {author} {\bibfnamefont {M.~B.}\ \bibnamefont
  {Plenio}}, \bibinfo {author} {\bibfnamefont {B.}~\bibnamefont {Naydenov}},
  \bibinfo {author} {\bibfnamefont {L.~P.}\ \bibnamefont {{McGuinness}}}, \
  and\ \bibinfo {author} {\bibfnamefont {F.}~\bibnamefont {Jelezko}},\
  }\bibfield  {title} {\enquote {\bibinfo {title} {Nuclear magnetic resonance
  spectroscopy with single spin sensitivity},}\ }\href@noop {} {\bibfield
  {journal} {\bibinfo  {journal} {Nat. Commun.}\ }\textbf {\bibinfo {volume}
  {5}},\ \bibinfo {pages} {4703} (\bibinfo {year} {2014})}\BibitemShut
  {NoStop}%
\bibitem [{\citenamefont {Cai}\ \emph {et~al.}(2013{\natexlab{a}})\citenamefont
  {Cai}, \citenamefont {Jelezko}, \citenamefont {Plenio},\ and\ \citenamefont
  {Retzker}}]{cai2013diamond}%
  \BibitemOpen
  \bibfield  {author} {\bibinfo {author} {\bibfnamefont {J.}~\bibnamefont
  {Cai}}, \bibinfo {author} {\bibfnamefont {F.}~\bibnamefont {Jelezko}},
  \bibinfo {author} {\bibfnamefont {M.~B.}\ \bibnamefont {Plenio}}, \ and\
  \bibinfo {author} {\bibfnamefont {A.}~\bibnamefont {Retzker}},\ }\bibfield
  {title} {\enquote {\bibinfo {title} {Diamond-based single-molecule magnetic
  resonance spectroscopy},}\ }\href@noop {} {\bibfield  {journal} {\bibinfo
  {journal} {New J. Phys.}\ }\textbf {\bibinfo {volume} {15}},\ \bibinfo
  {pages} {013020} (\bibinfo {year} {2013}{\natexlab{a}})}\BibitemShut
  {NoStop}%
\bibitem [{\citenamefont {Kost}\ \emph {et~al.}(2015)\citenamefont {Kost},
  \citenamefont {Cai},\ and\ \citenamefont {Plenio}}]{kost2014resolving}%
  \BibitemOpen
  \bibfield  {author} {\bibinfo {author} {\bibfnamefont {M.}~\bibnamefont
  {Kost}}, \bibinfo {author} {\bibfnamefont {J.}~\bibnamefont {Cai}}, \ and\
  \bibinfo {author} {\bibfnamefont {M.~B.}\ \bibnamefont {Plenio}},\ }\bibfield
   {title} {\enquote {\bibinfo {title} {Resolving single molecule structures
  with nitrogen-vacancy centers in diamond},}\ }\href@noop {} {\bibfield
  {journal} {\bibinfo  {journal} {Sci. Rep.}\ }\textbf {\bibinfo {volume}
  {5}},\ \bibinfo {pages} {11007} (\bibinfo {year} {2015})}\BibitemShut
  {NoStop}%
\bibitem [{\citenamefont {Ajoy}\ \emph {et~al.}(2015)\citenamefont {Ajoy},
  \citenamefont {Bissbort}, \citenamefont {Lukin}, \citenamefont {Walsworth},\
  and\ \citenamefont {Cappellaro}}]{ajoy2015atomic}%
  \BibitemOpen
  \bibfield  {author} {\bibinfo {author} {\bibfnamefont {A.}~\bibnamefont
  {Ajoy}}, \bibinfo {author} {\bibfnamefont {U.}~\bibnamefont {Bissbort}},
  \bibinfo {author} {\bibfnamefont {M.~D.}\ \bibnamefont {Lukin}}, \bibinfo
  {author} {\bibfnamefont {R.~L.}\ \bibnamefont {Walsworth}}, \ and\ \bibinfo
  {author} {\bibfnamefont {P.}~\bibnamefont {Cappellaro}},\ }\bibfield  {title}
  {\enquote {\bibinfo {title} {Atomic-scale nuclear spin imaging using
  quantum-assisted sensors in diamond},}\ }\href@noop {} {\bibfield  {journal}
  {\bibinfo  {journal} {Phys. Rev. X}\ }\textbf {\bibinfo {volume} {5}},\
  \bibinfo {pages} {011001} (\bibinfo {year} {2015})}\BibitemShut {NoStop}%
\bibitem [{\citenamefont {Laraoui}\ \emph {et~al.}(2015)\citenamefont
  {Laraoui}, \citenamefont {Pagliero},\ and\ \citenamefont
  {Meriles}}]{laraoui2015imaging}%
  \BibitemOpen
  \bibfield  {author} {\bibinfo {author} {\bibfnamefont {A.}~\bibnamefont
  {Laraoui}}, \bibinfo {author} {\bibfnamefont {D.}~\bibnamefont {Pagliero}}, \
  and\ \bibinfo {author} {\bibfnamefont {C.~A.}\ \bibnamefont {Meriles}},\
  }\bibfield  {title} {\enquote {\bibinfo {title} {Imaging nuclear spins weakly
  coupled to a probe paramagnetic center},}\ }\href@noop {} {\bibfield
  {journal} {\bibinfo  {journal} {Phys. Rev. B}\ }\textbf {\bibinfo {volume}
  {91}},\ \bibinfo {pages} {205410} (\bibinfo {year} {2015})}\BibitemShut
  {NoStop}%
\bibitem [{\citenamefont {Epstein}\ \emph {et~al.}(2005)\citenamefont
  {Epstein}, \citenamefont {Mendoza}, \citenamefont {Kato},\ and\ \citenamefont
  {Awschalom}}]{Epstein:2005:94}%
  \BibitemOpen
  \bibfield  {author} {\bibinfo {author} {\bibfnamefont {R.~J.}\ \bibnamefont
  {Epstein}}, \bibinfo {author} {\bibfnamefont {F.~M.}\ \bibnamefont
  {Mendoza}}, \bibinfo {author} {\bibfnamefont {Y.~K.}\ \bibnamefont {Kato}}, \
  and\ \bibinfo {author} {\bibfnamefont {D.~D.}\ \bibnamefont {Awschalom}},\
  }\bibfield  {title} {\enquote {\bibinfo {title} {Anisotropic interactions of
  a single spin and dark-spin spectroscopy in diamond},}\ }\href@noop {}
  {\bibfield  {journal} {\bibinfo  {journal} {Nat. Phys.}\ }\textbf {\bibinfo
  {volume} {1}},\ \bibinfo {pages} {94} (\bibinfo {year} {2005})}\BibitemShut
  {NoStop}%
\bibitem [{\citenamefont {Lee}\ and\ \citenamefont
  {Goldburg}(1965)}]{lee1965nuclear}%
  \BibitemOpen
  \bibfield  {author} {\bibinfo {author} {\bibfnamefont {M.}~\bibnamefont
  {Lee}}\ and\ \bibinfo {author} {\bibfnamefont {W.~I.}\ \bibnamefont
  {Goldburg}},\ }\bibfield  {title} {\enquote {\bibinfo {title}
  {Nuclear-magnetic-resonance line narrowing by a rotating rf field},}\
  }\href@noop {} {\bibfield  {journal} {\bibinfo  {journal} {Phys. Rev.}\
  }\textbf {\bibinfo {volume} {140}},\ \bibinfo {pages} {A1261} (\bibinfo
  {year} {1965})}\BibitemShut {NoStop}%
\bibitem [{\citenamefont {Bielecki}\ \emph {et~al.}(1989)\citenamefont
  {Bielecki}, \citenamefont {Kolbert},\ and\ \citenamefont
  {Levitt}}]{bielecki1989frequency}%
  \BibitemOpen
  \bibfield  {author} {\bibinfo {author} {\bibfnamefont {A.}~\bibnamefont
  {Bielecki}}, \bibinfo {author} {\bibfnamefont {A.~C.}\ \bibnamefont
  {Kolbert}}, \ and\ \bibinfo {author} {\bibfnamefont {M.~H.}\ \bibnamefont
  {Levitt}},\ }\bibfield  {title} {\enquote {\bibinfo {title}
  {Frequency-switched pulse sequences: homonuclear decoupling and dilute spin
  {NMR} in solids},}\ }\href@noop {} {\bibfield  {journal} {\bibinfo  {journal}
  {Chem. Phys. Lett.}\ }\textbf {\bibinfo {volume} {155}},\ \bibinfo {pages}
  {341} (\bibinfo {year} {1989})}\BibitemShut {NoStop}%
\bibitem [{\citenamefont {Vinogradov}\ \emph {et~al.}(2002)\citenamefont
  {Vinogradov}, \citenamefont {Madhu},\ and\ \citenamefont
  {Vega}}]{vinogradov2002proton}%
  \BibitemOpen
  \bibfield  {author} {\bibinfo {author} {\bibfnamefont {E.}~\bibnamefont
  {Vinogradov}}, \bibinfo {author} {\bibfnamefont {P.~K.}\ \bibnamefont
  {Madhu}}, \ and\ \bibinfo {author} {\bibfnamefont {S.}~\bibnamefont {Vega}},\
  }\bibfield  {title} {\enquote {\bibinfo {title} {Proton spectroscopy in solid
  state nuclear magnetic resonance with windowed phase modulated {Lee-Goldburg}
  decoupling sequences},}\ }\href@noop {} {\bibfield  {journal} {\bibinfo
  {journal} {Chem. Phys. Lett.}\ }\textbf {\bibinfo {volume} {354}},\ \bibinfo
  {pages} {193} (\bibinfo {year} {2002})}\BibitemShut {NoStop}%
\bibitem [{\citenamefont {Cai}\ \emph {et~al.}(2013{\natexlab{b}})\citenamefont
  {Cai}, \citenamefont {Retzker}, \citenamefont {Jelezko},\ and\ \citenamefont
  {Plenio}}]{cai2013large}%
  \BibitemOpen
  \bibfield  {author} {\bibinfo {author} {\bibfnamefont {J.}~\bibnamefont
  {Cai}}, \bibinfo {author} {\bibfnamefont {A.}~\bibnamefont {Retzker}},
  \bibinfo {author} {\bibfnamefont {F.}~\bibnamefont {Jelezko}}, \ and\
  \bibinfo {author} {\bibfnamefont {M.~B.}\ \bibnamefont {Plenio}},\ }\bibfield
   {title} {\enquote {\bibinfo {title} {A large-scale quantum simulator on a
  diamond surface at room temperature},}\ }\href@noop {} {\bibfield  {journal}
  {\bibinfo  {journal} {Nat. Phys.}\ }\textbf {\bibinfo {volume} {9}},\
  \bibinfo {pages} {168} (\bibinfo {year} {2013}{\natexlab{b}})}\BibitemShut
  {NoStop}%
\bibitem [{\citenamefont {Wang}\ \emph {et~al.}(2014)\citenamefont {Wang},
  \citenamefont {Cai}, \citenamefont {Retzker},\ and\ \citenamefont
  {Plenio}}]{wang2014all}%
  \BibitemOpen
  \bibfield  {author} {\bibinfo {author} {\bibfnamefont {Z.-Y.}\ \bibnamefont
  {Wang}}, \bibinfo {author} {\bibfnamefont {J.}~\bibnamefont {Cai}}, \bibinfo
  {author} {\bibfnamefont {A.}~\bibnamefont {Retzker}}, \ and\ \bibinfo
  {author} {\bibfnamefont {M.~B.}\ \bibnamefont {Plenio}},\ }\bibfield  {title}
  {\enquote {\bibinfo {title} {All-optical magnetic resonance of high spectral
  resolution using a nitrogen-vacancy spin in diamond},}\ }\href@noop {}
  {\bibfield  {journal} {\bibinfo  {journal} {New J. Phys.}\ }\textbf {\bibinfo
  {volume} {16}},\ \bibinfo {pages} {083033} (\bibinfo {year}
  {2014})}\BibitemShut {NoStop}%
\bibitem [{\citenamefont {Yang}\ \emph {et~al.}(2011)\citenamefont {Yang},
  \citenamefont {Wang},\ and\ \citenamefont {Liu}}]{yang2011preserving}%
  \BibitemOpen
  \bibfield  {author} {\bibinfo {author} {\bibfnamefont {W.}~\bibnamefont
  {Yang}}, \bibinfo {author} {\bibfnamefont {Z.-Y.}\ \bibnamefont {Wang}}, \
  and\ \bibinfo {author} {\bibfnamefont {R.-B.}\ \bibnamefont {Liu}},\
  }\bibfield  {title} {\enquote {\bibinfo {title} {Preserving qubit coherence
  by dynamical decoupling},}\ }\href@noop {} {\bibfield  {journal} {\bibinfo
  {journal} {Front. Phys.}\ }\textbf {\bibinfo {volume} {6}},\ \bibinfo {pages}
  {2} (\bibinfo {year} {2011})}\BibitemShut {NoStop}%
\bibitem [{\citenamefont {Carr}\ and\ \citenamefont
  {Purcell}(1954)}]{Carr:1954:630}%
  \BibitemOpen
  \bibfield  {author} {\bibinfo {author} {\bibfnamefont {H.~Y.}\ \bibnamefont
  {Carr}}\ and\ \bibinfo {author} {\bibfnamefont {E.~M.}\ \bibnamefont
  {Purcell}},\ }\bibfield  {title} {\enquote {\bibinfo {title} {Effects of
  diffusion on free precession in nuclear magnetic resonance experiments},}\
  }\href@noop {} {\bibfield  {journal} {\bibinfo  {journal} {Phys. Rev.}\
  }\textbf {\bibinfo {volume} {94}},\ \bibinfo {pages} {630} (\bibinfo {year}
  {1954})}\BibitemShut {NoStop}%
\bibitem [{\citenamefont {Meiboom}\ and\ \citenamefont
  {Gill}(1958)}]{MeiboomRSI1958}%
  \BibitemOpen
  \bibfield  {author} {\bibinfo {author} {\bibfnamefont {S.}~\bibnamefont
  {Meiboom}}\ and\ \bibinfo {author} {\bibfnamefont {D.}~\bibnamefont {Gill}},\
  }\bibfield  {title} {\enquote {\bibinfo {title} {Modified spin-echo method
  for measuring nuclear relaxation times},}\ }\href@noop {} {\bibfield
  {journal} {\bibinfo  {journal} {Rev. Sci. Instrum.}\ }\textbf {\bibinfo
  {volume} {29}},\ \bibinfo {pages} {688} (\bibinfo {year} {1958})}\BibitemShut
  {NoStop}%
\bibitem [{\citenamefont {Maudsley}(1986)}]{MaudsleyJMR1986}%
  \BibitemOpen
  \bibfield  {author} {\bibinfo {author} {\bibfnamefont {A.~A.}\ \bibnamefont
  {Maudsley}},\ }\bibfield  {title} {\enquote {\bibinfo {title} {Modified
  {Carr-Purcell-Meiboom-Gill} sequence for {NMR} fourier imaging
  applications},}\ }\href@noop {} {\bibfield  {journal} {\bibinfo  {journal}
  {J. Magn. Reson. (1969)}\ }\textbf {\bibinfo {volume} {69}},\ \bibinfo
  {pages} {488} (\bibinfo {year} {1986})}\BibitemShut {NoStop}%
\bibitem [{\citenamefont {Gullion}\ \emph {et~al.}(1990)\citenamefont
  {Gullion}, \citenamefont {Baker},\ and\ \citenamefont
  {Conradi}}]{GullionJMR1990}%
  \BibitemOpen
  \bibfield  {author} {\bibinfo {author} {\bibfnamefont {T.}~\bibnamefont
  {Gullion}}, \bibinfo {author} {\bibfnamefont {D.~B.}\ \bibnamefont {Baker}},
  \ and\ \bibinfo {author} {\bibfnamefont {M.~S.}\ \bibnamefont {Conradi}},\
  }\bibfield  {title} {\enquote {\bibinfo {title} {New, compensated
  {Carr-Purcell} sequences},}\ }\href@noop {} {\bibfield  {journal} {\bibinfo
  {journal} {J. Magn. Reson. (1969)}\ }\textbf {\bibinfo {volume} {89}},\
  \bibinfo {pages} {479} (\bibinfo {year} {1990})}\BibitemShut {NoStop}%
\bibitem [{\citenamefont {Zhao}\ \emph {et~al.}(2014)\citenamefont {Zhao},
  \citenamefont {Wrachtrup},\ and\ \citenamefont {Liu}}]{Zhao2014KDD}%
  \BibitemOpen
  \bibfield  {author} {\bibinfo {author} {\bibfnamefont {N.}~\bibnamefont
  {Zhao}}, \bibinfo {author} {\bibfnamefont {J.}~\bibnamefont {Wrachtrup}}, \
  and\ \bibinfo {author} {\bibfnamefont {R.-B.}\ \bibnamefont {Liu}},\
  }\bibfield  {title} {\enquote {\bibinfo {title} {Dynamical decoupling design
  for identifying weakly coupled nuclear spins in a bath},}\ }\href@noop {}
  {\bibfield  {journal} {\bibinfo  {journal} {Phys. Rev. A}\ }\textbf {\bibinfo
  {volume} {90}},\ \bibinfo {pages} {032319} (\bibinfo {year}
  {2014})}\BibitemShut {NoStop}%
\bibitem [{\citenamefont {Albrecht}\ and\ \citenamefont
  {Plenio}(2015)}]{AlbrechtP15}%
  \BibitemOpen
  \bibfield  {author} {\bibinfo {author} {\bibfnamefont {A.}~\bibnamefont
  {Albrecht}}\ and\ \bibinfo {author} {\bibfnamefont {M.~B.}\ \bibnamefont
  {Plenio}},\ }\bibfield  {title} {\enquote {\bibinfo {title} {Filter design
  for hybrid spin gates},}\ }\href@noop {} {\bibfield  {journal} {\bibinfo
  {journal} {Phys. Rev. A}\ }\textbf {\bibinfo {volume} {92}},\ \bibinfo
  {pages} {022340} (\bibinfo {year} {2015})}\BibitemShut {NoStop}%
\bibitem [{\citenamefont {Ma}\ \emph {et~al.}(2015)\citenamefont {Ma},
  \citenamefont {Shi}, \citenamefont {Xu}, \citenamefont {Wang}, \citenamefont
  {Xu}, \citenamefont {Rong}, \citenamefont {Ju}, \citenamefont {Duan},
  \citenamefont {Zhao},\ and\ \citenamefont {Du}}]{Ma2015resolving}%
  \BibitemOpen
  \bibfield  {author} {\bibinfo {author} {\bibfnamefont {W.}~\bibnamefont
  {Ma}}, \bibinfo {author} {\bibfnamefont {F.}~\bibnamefont {Shi}}, \bibinfo
  {author} {\bibfnamefont {K.}~\bibnamefont {Xu}}, \bibinfo {author}
  {\bibfnamefont {P.}~\bibnamefont {Wang}}, \bibinfo {author} {\bibfnamefont
  {X.}~\bibnamefont {Xu}}, \bibinfo {author} {\bibfnamefont {X.}~\bibnamefont
  {Rong}}, \bibinfo {author} {\bibfnamefont {C.}~\bibnamefont {Ju}}, \bibinfo
  {author} {\bibfnamefont {C.-K.}\ \bibnamefont {Duan}}, \bibinfo {author}
  {\bibfnamefont {N.}~\bibnamefont {Zhao}}, \ and\ \bibinfo {author}
  {\bibfnamefont {J.}~\bibnamefont {Du}},\ }\bibfield  {title} {\enquote
  {\bibinfo {title} {Resolving remote nuclear spins in a noisy bath by
  dynamical decoupling design},}\ }\href@noop {} {\bibfield  {journal}
  {\bibinfo  {journal} {Phys. Rev. A}\ }\textbf {\bibinfo {volume} {92}},\
  \bibinfo {pages} {033418} (\bibinfo {year} {2015})}\BibitemShut {NoStop}%
\bibitem [{\citenamefont {Maze}\ \emph {et~al.}(2008)\citenamefont {Maze},
  \citenamefont {Taylor},\ and\ \citenamefont {Lukin}}]{Maze2008PRB}%
  \BibitemOpen
  \bibfield  {author} {\bibinfo {author} {\bibfnamefont {J.~R.}\ \bibnamefont
  {Maze}}, \bibinfo {author} {\bibfnamefont {J.~M.}\ \bibnamefont {Taylor}}, \
  and\ \bibinfo {author} {\bibfnamefont {M.~D.}\ \bibnamefont {Lukin}},\
  }\bibfield  {title} {\enquote {\bibinfo {title} {Electron spin decoherence of
  single nitrogen-vacancy defects in diamond},}\ }\href@noop {} {\bibfield
  {journal} {\bibinfo  {journal} {Phys. Rev. B}\ }\textbf {\bibinfo {volume}
  {78}},\ \bibinfo {pages} {094303} (\bibinfo {year} {2008})}\BibitemShut
  {NoStop}%
\bibitem [{\citenamefont {Yang}\ and\ \citenamefont {Liu}(2008)}]{Yang2008PRB}%
  \BibitemOpen
  \bibfield  {author} {\bibinfo {author} {\bibfnamefont {W.}~\bibnamefont
  {Yang}}\ and\ \bibinfo {author} {\bibfnamefont {R.-B.}\ \bibnamefont {Liu}},\
  }\bibfield  {title} {\enquote {\bibinfo {title} {Quantum many-body theory of
  qubit decoherence in a finite-size spin bath},}\ }\href@noop {} {\bibfield
  {journal} {\bibinfo  {journal} {Phys. Rev. B}\ }\textbf {\bibinfo {volume}
  {78}},\ \bibinfo {pages} {085315} (\bibinfo {year} {2008})}\BibitemShut
  {NoStop}%
\bibitem [{\citenamefont {Loretz}\ \emph {et~al.}(2015)\citenamefont {Loretz},
  \citenamefont {Boss}, \citenamefont {Rosskopf}, \citenamefont {Mamin},
  \citenamefont {Rugar},\ and\ \citenamefont {Degen}}]{loretz2015spurious}%
  \BibitemOpen
  \bibfield  {author} {\bibinfo {author} {\bibfnamefont {M.}~\bibnamefont
  {Loretz}}, \bibinfo {author} {\bibfnamefont {J.~M.}\ \bibnamefont {Boss}},
  \bibinfo {author} {\bibfnamefont {T.}~\bibnamefont {Rosskopf}}, \bibinfo
  {author} {\bibfnamefont {H.~J.}\ \bibnamefont {Mamin}}, \bibinfo {author}
  {\bibfnamefont {D.}~\bibnamefont {Rugar}}, \ and\ \bibinfo {author}
  {\bibfnamefont {C.~L.}\ \bibnamefont {Degen}},\ }\bibfield  {title} {\enquote
  {\bibinfo {title} {Spurious harmonic response of multipulse quantum sensing
  sequences},}\ }\href@noop {} {\bibfield  {journal} {\bibinfo  {journal}
  {Phys. Rev. X}\ }\textbf {\bibinfo {volume} {5}},\ \bibinfo {pages} {021009}
  (\bibinfo {year} {2015})}\BibitemShut {NoStop}%
\bibitem [{\citenamefont {Michal}\ \emph {et~al.}(2008)\citenamefont {Michal},
  \citenamefont {Hastings},\ and\ \citenamefont {Lee}}]{michal2008two}%
  \BibitemOpen
  \bibfield  {author} {\bibinfo {author} {\bibfnamefont {C.~A.}\ \bibnamefont
  {Michal}}, \bibinfo {author} {\bibfnamefont {S.~P.}\ \bibnamefont
  {Hastings}}, \ and\ \bibinfo {author} {\bibfnamefont {L.~H.}\ \bibnamefont
  {Lee}},\ }\bibfield  {title} {\enquote {\bibinfo {title} {Two-photon
  {Lee-Goldburg} nuclear magnetic resonance: Simultaneous homonuclear
  decoupling and signal acquisition},}\ }\href@noop {} {\bibfield  {journal}
  {\bibinfo  {journal} {J. Chem. Phys.}\ }\textbf {\bibinfo {volume} {128}},\
  \bibinfo {pages} {052301} (\bibinfo {year} {2008})}\BibitemShut {NoStop}%
\bibitem [{\citenamefont {Fuchs}\ \emph {et~al.}(2009)\citenamefont {Fuchs},
  \citenamefont {Dobrovitski}, \citenamefont {Toyli}, \citenamefont
  {Heremans},\ and\ \citenamefont {Awschalom}}]{fuchs2009gigahertz}%
  \BibitemOpen
  \bibfield  {author} {\bibinfo {author} {\bibfnamefont {G.~D.}\ \bibnamefont
  {Fuchs}}, \bibinfo {author} {\bibfnamefont {V.~V.}\ \bibnamefont
  {Dobrovitski}}, \bibinfo {author} {\bibfnamefont {D.~M.}\ \bibnamefont
  {Toyli}}, \bibinfo {author} {\bibfnamefont {F.~J.}\ \bibnamefont {Heremans}},
  \ and\ \bibinfo {author} {\bibfnamefont {D.~D.}\ \bibnamefont {Awschalom}},\
  }\bibfield  {title} {\enquote {\bibinfo {title} {Gigahertz dynamics of a
  strongly driven single quantum spin},}\ }\href@noop {} {\bibfield  {journal}
  {\bibinfo  {journal} {Science}\ }\textbf {\bibinfo {volume} {326}},\ \bibinfo
  {pages} {1520} (\bibinfo {year} {2009})}\BibitemShut {NoStop}%
\bibitem [{\citenamefont {Abragam}(1961)}]{abragam1961principles}%
  \BibitemOpen
  \bibfield  {author} {\bibinfo {author} {\bibfnamefont {A.}~\bibnamefont
  {Abragam}},\ }\href@noop {} {\emph {\bibinfo {title} {The principles of
  nuclear magnetism}}},\ \bibinfo {number} {32}\ (\bibinfo  {publisher} {Oxford
  university press},\ \bibinfo {year} {1961})\BibitemShut {NoStop}%
\bibitem [{\citenamefont {Romach}\ \emph {et~al.}(2015)\citenamefont {Romach},
  \citenamefont {M\"uller}, \citenamefont {Unden}, \citenamefont {Rogers},
  \citenamefont {Isoda}, \citenamefont {Itoh}, \citenamefont {Markham},
  \citenamefont {Stacey}, \citenamefont {Meijer}, \citenamefont {Pezzagna},
  \citenamefont {Naydenov}, \citenamefont {{McGuinness}}, \citenamefont
  {{Bar-Gill}},\ and\ \citenamefont {Jelezko}}]{romach2015spectroscopy}%
  \BibitemOpen
  \bibfield  {author} {\bibinfo {author} {\bibfnamefont {Y.}~\bibnamefont
  {Romach}}, \bibinfo {author} {\bibfnamefont {C.}~\bibnamefont {M\"uller}},
  \bibinfo {author} {\bibfnamefont {T.}~\bibnamefont {Unden}}, \bibinfo
  {author} {\bibfnamefont {L.~J.}\ \bibnamefont {Rogers}}, \bibinfo {author}
  {\bibfnamefont {T.}~\bibnamefont {Isoda}}, \bibinfo {author} {\bibfnamefont
  {K.~M.}\ \bibnamefont {Itoh}}, \bibinfo {author} {\bibfnamefont
  {M.}~\bibnamefont {Markham}}, \bibinfo {author} {\bibfnamefont
  {A.}~\bibnamefont {Stacey}}, \bibinfo {author} {\bibfnamefont
  {J.}~\bibnamefont {Meijer}}, \bibinfo {author} {\bibfnamefont
  {S.}~\bibnamefont {Pezzagna}}, \bibinfo {author} {\bibfnamefont
  {B.}~\bibnamefont {Naydenov}}, \bibinfo {author} {\bibfnamefont {L.~P.}\
  \bibnamefont {{McGuinness}}}, \bibinfo {author} {\bibfnamefont
  {N.}~\bibnamefont {{Bar-Gill}}}, \ and\ \bibinfo {author} {\bibfnamefont
  {F.}~\bibnamefont {Jelezko}},\ }\bibfield  {title} {\enquote {\bibinfo
  {title} {Spectroscopy of surface-induced noise using shallow spins in
  diamond},}\ }\href@noop {} {\bibfield  {journal} {\bibinfo  {journal} {Phys.
  Rev. Lett.}\ }\textbf {\bibinfo {volume} {114}},\ \bibinfo {pages} {017601}
  (\bibinfo {year} {2015})}\BibitemShut {NoStop}%
\bibitem [{\citenamefont {Lovchinsky}\ \emph {et~al.}(2016)\citenamefont
  {Lovchinsky}, \citenamefont {Sushkov}, \citenamefont {Urbach}, \citenamefont
  {de~Leon}, \citenamefont {Choi}, \citenamefont {De~Greve}, \citenamefont
  {Evans}, \citenamefont {Gertner}, \citenamefont {Bersin}, \citenamefont
  {M{\"u}ller}, \citenamefont {{McGuinness}}, \citenamefont {Jelezko},
  \citenamefont {Walsworth}, \citenamefont {Park},\ and\ \citenamefont
  {Lukin}}]{lovchinsky2016nuclear}%
  \BibitemOpen
  \bibfield  {author} {\bibinfo {author} {\bibfnamefont {I.}~\bibnamefont
  {Lovchinsky}}, \bibinfo {author} {\bibfnamefont {A.~O.}\ \bibnamefont
  {Sushkov}}, \bibinfo {author} {\bibfnamefont {E.}~\bibnamefont {Urbach}},
  \bibinfo {author} {\bibfnamefont {N.~P.}\ \bibnamefont {de~Leon}}, \bibinfo
  {author} {\bibfnamefont {S.}~\bibnamefont {Choi}}, \bibinfo {author}
  {\bibfnamefont {K.}~\bibnamefont {De~Greve}}, \bibinfo {author}
  {\bibfnamefont {R.}~\bibnamefont {Evans}}, \bibinfo {author} {\bibfnamefont
  {R.}~\bibnamefont {Gertner}}, \bibinfo {author} {\bibfnamefont
  {E.}~\bibnamefont {Bersin}}, \bibinfo {author} {\bibfnamefont
  {C.}~\bibnamefont {M{\"u}ller}}, \bibinfo {author} {\bibfnamefont
  {L.}~\bibnamefont {{McGuinness}}}, \bibinfo {author} {\bibfnamefont
  {F.}~\bibnamefont {Jelezko}}, \bibinfo {author} {\bibfnamefont {R.~L.}\
  \bibnamefont {Walsworth}}, \bibinfo {author} {\bibfnamefont {H.}~\bibnamefont
  {Park}}, \ and\ \bibinfo {author} {\bibfnamefont {M.~D.}\ \bibnamefont
  {Lukin}},\ }\bibfield  {title} {\enquote {\bibinfo {title} {Nuclear magnetic
  resonance detection and spectroscopy of single proteins using quantum
  logic},}\ }\href@noop {} {\bibfield  {journal} {\bibinfo  {journal}
  {Science}\ }\textbf {\bibinfo {volume} {351}},\ \bibinfo {pages} {836}
  (\bibinfo {year} {2016})}\BibitemShut {NoStop}%
\end{thebibliography}
%

\end{document}